\documentclass[a4paper,11pt]{article}
\usepackage{jheppub}

\usepackage{chessfss}

\usepackage{hyperref}
\hypersetup{colorlinks=true,linkcolor=blue,citecolor=blue}
\usepackage{amsmath,amssymb,amsfonts}
\usepackage{bm}
\usepackage{float}
\usepackage{subfig}
\usepackage{scalefnt}
\usepackage{wrapfig}
\usepackage{todonotes}
\usepackage{fancybox}
\usepackage{graphicx}
\usepackage{tikz}
\usepackage{multirow}
\usepackage{adjustbox}
\usepackage[toc,page]{appendix}

\allowdisplaybreaks

\usetikzlibrary{arrows}
\tikzset{>=angle 60}
\tikzstyle{W}=[draw,circle,scale=.6]
\tikzstyle{B}=[draw,circle,fill=black,scale=.6]
\tikzstyle{H}=[draw,circle,fill=gray,scale=.6]
\tikzstyle{every picture}=[scale=.6,baseline=(current bounding box.south)]

\def\F{\Phi}

\def\beq#1\eeq{\begin{align}#1\end{align}}

\newcommand{\be}{\begin{eqnarray}}
\newcommand{\ee}{\end{eqnarray}}
\newcommand{\bea}{\begin{eqnarray}}
\newcommand{\eea}{\end{eqnarray}}

\newcommand{\bn}{\begin{enumerate}}
\newcommand{\en}{\end{enumerate}}

\def\IC{\mathbb{C}}

\def\IR{\mathbb{R}}
\def\IZ{\mathbb{Z}}

\def\i{\iota}

\def\half{\frac{1}{2}}

\newcommand{\ket}[1]{|{#1}\rangle}

\usepackage{mathtools,amsmath}
\usepackage{verbatim}
\newcommand{\upa}{\uparrow}
\newcommand{\doa}{\downarrow}

\title{Ising-like and Fibonacci Anyons from KZ-equations}

\begin{document}
\author{Xia Gu,}
\author{Babak Haghighat,}
\author{Yihua Liu}
\affiliation{Yau Mathematical Sciences Center, Tsinghua University, Beijing, 100084, China}
\emailAdd{gux19@mails.tsinghua.edu.cn, babakhaghighat@tsinghua.edu.cn, liuyh19@mails.tsinghua.edu.cn}

\abstract{In this work we present solutions to Knizhnik-Zamolodchikov (KZ) equations corresponding to conformal block wavefunctions of non-Abelian Ising-like and Fibonacci Anyons. We solve these equations around regular singular points in configuration space in terms of hypergeometric functions and derive explicit monodromy representations of the braid group action. This confirms the correct non-Abelian statistics of the solutions. One novelty of our approach is that we explicitly keep track of spin basis states and identify conformal blocks uniquely with such states at relevant points in moduli space.}

\maketitle

\section{Introduction}

The Knizhnik-Zamolodchikov (KZ) equations \cite{KNIZHNIK198483} describe a flat vector bundle, the so-called \textit{conformal block bundle}, over the configuration space of $n$ marked points on the sphere. They are formulated by using the KZ connection to define a horizontal section of such a bundle. In more physical terms such a section can be viewed as a wavefunction and the flatness of the connection just implies that parallel transport of the section along the configuration space does not depend on the path chosen. This way the connection can be used to define the wavefunction consistently everywhere. 

Now one might ask what quantum mechanical system these wavefunctions belong to. This question was answered by Witten \cite{Witten:1988hf,Witten:1991mm} by viewing the 2-sphere (or any Riemann surface for that matter) as the boundary of a 3-manifold and the KZ wavefunction as a quantum state of Chern-Simons theory on that 3-manifold with Wilson lines ending on the 2-sphere at $n$ points. For our purposes, the 3-ball is more than sufficient to present the picture we want to convey. The corresponding wavefunction $\psi$ then takes values in the $n$-fold tensor product of Wilson line representations such that the outcome is a singlet of the Chern-Simons gauge group. Typically there are several such singlets and the wavefunction is a linear combination of them. 

Another interpretation of conformal blocks as wavefunctions goes back to the work of Moore and Read \cite{Moore:1991ks} where the idea is to use these as quantum states of topological phases of matter. Within this correspondence, the wavefunctions become eigenstates of a Hamiltonian describing a quantum Hall or lattice spin system. Here, the chiral field insertions in the conformal block correspond to Anyons, i.e. particles with fractional statistics, which appear as excitations of the system. One subdivides then between Abelian and non-Abelian phases. In the Abelian phase the exchange of two Anyons produces just an abelian phase, while in the non-Abelian case the energy eigenstates of the Hamiltonian are degenerate and exchange of two Anyons produces non-commuting matrices acting on this degenerate subspace, see \cite{Nayak_2008} for a nice review on this subject. The relevant phase for us will be the non-Abelian phase and the argument of \cite{Moore:1991ks} to relate its quantum states to conformal blocks goes in a modern setting roughly as follows. The two-dimensional space where the Anyons live can be viewed as a fixed time slice of a $(2+1)$-dimensional theory which is in fact gapped. This is because the spectrum of the Hamiltonian separates excitations from the degenerate subspace by an energy gap as for example shown explicitly in \cite{Kitaev_2006} for the case of Ising Anyons. Thus the bulk theory can be viewed as a $(2+1)$-d TQFT with the special property that its edge excitations on the boundary are gapless chiral modes. This leads to the chiral CFT formulation of the boundary theory which ultimately gives rise to the proposal of Moore and Read. Chern-Simons theory is an instance of such a TQFT which connects the present discussion to the paragraph before\footnote{When one is dealing with spin systems, it is more natural to view the bulk theory as a lattice TQFT such that the degrees of freedom are situated on vertices and links.}. The Anyons of the boundary theory extend to line defects inside the bulk which can braid and form trivalent junctions, as shown in the picture below.

\begin{figure}[h]
  \centering
	\includegraphics[width=0.4\textwidth]{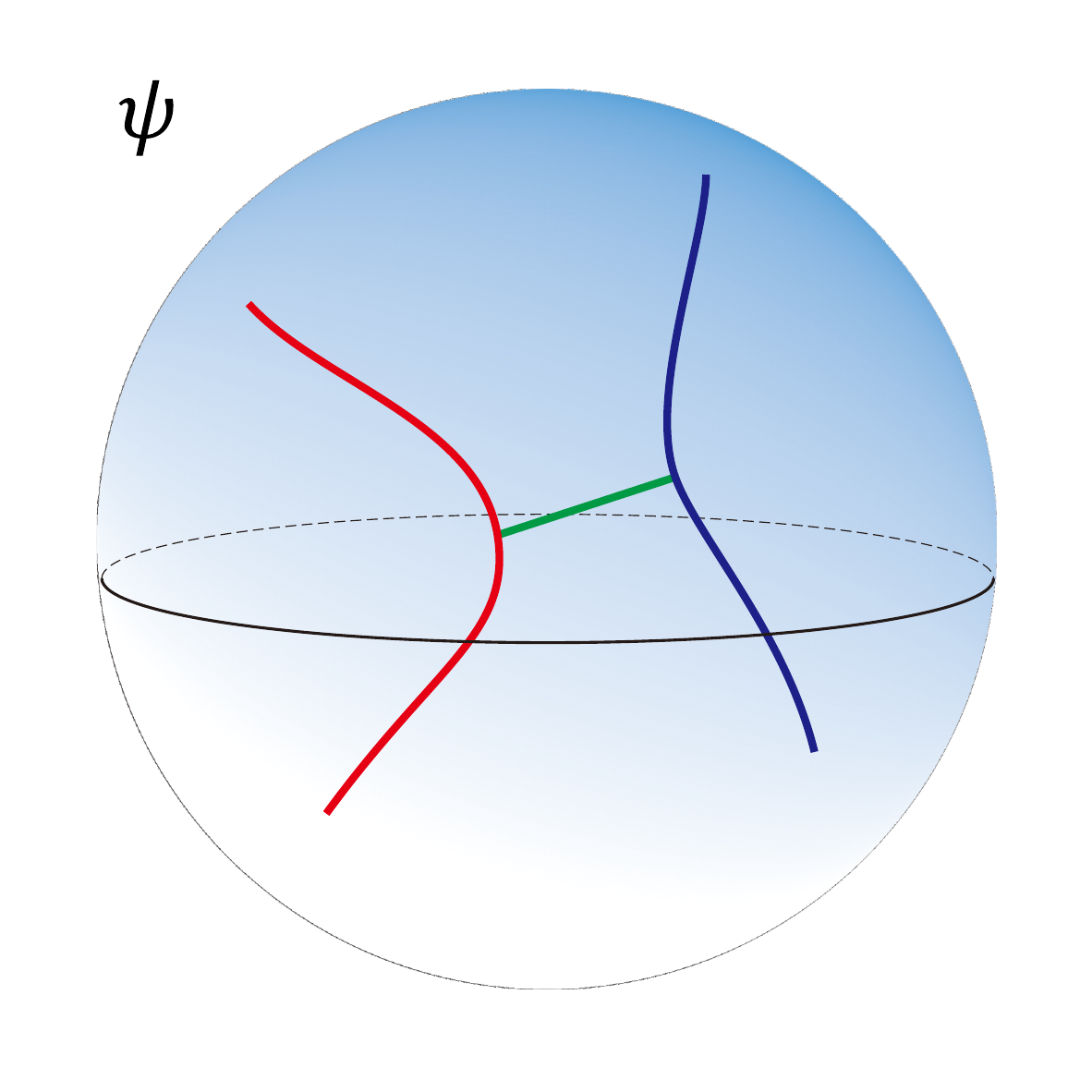}
  \caption{Anyons as line defects in the bulk. $\psi$ denotes the wavefunction corresponding to the boundary state of the defect configuration.}
  \label{fig:buld-boundary}
\end{figure}

These types of non-Abelian anyons are closely related to ``non-invertible" defects in the more recent literature where new and surprising properties are currently being investigated \cite{Bhardwaj:2018,Chang:2019,Kaidi:2021,Choi:2021kmx,Komargodski:2020mxz,Kaidi:2021xfk} (see also \cite{Aasen:2020jwb} for a discussion in the case of lattice models). 

As already mentioned, one way to solve for conformal block wavefunctions is to solve the KZ equations which are formulated in terms of a flat connection. In other words, the Berry phase is zero along any paths in the configuration manifold. For the Laughlin states, the Berry phase is explicitly shown to be zero in \cite{PhysRevLett.53.722}. And the Ising case is treated in \cite{Bonderson_2011}. The connection being flat doesn't, however, mean that transport of solutions from one point to another is completely path independent as there could be monodromy. And indeed this is what happens and solutions are acted upon by a non-trivial monodromy matrix when transported around singularities in moduli space. This precisely accounts for the non-Abelian statistics of Anyon exchange. In the current work we solve the KZ-equations for non-Abelian Ising-like and Fibonacci Anyons by viewing these as subalgebras of chiral primary sectors of $\widehat{\mathfrak{su}}(2)_2$ and $\widehat{\mathfrak{su}}(2)_3$ WZW models. We then check that our solutions have the correct statistics by an explicit computation of the monodromy representation for the case of 4-point and 6-points functions and argue that higher-point cases can be reduced to these. Chiral conformal blocks for Ising Anyons had been previously computed in the literature \cite{Nayak_1996}, see also \cite{Fendley:1995xa,Ardonne:2010hj}. The Fibonacci anyons were also considered by Ardonne-Schoutens \cite{Ardonne_2007}. One novelty of our approach as compared to these previous approaches is that we explicitly keep track of spin states in the solution and identify different solutions with unique spin basis states at singular points in moduli space. Although non-Abelian Anyons are believed to not carry any intrinsic degrees of freedom apart from their association to simple objects in a fusion category, ground state wavefunctions of topological phases had been constructed previously using explicit spin basis conformal blocks \cite{Dorey:2016mxm,Zhang:2020dww} and thus it is conceivable that such explicit spin information will be important for the construction of excited states along the same lines. 

The organization of this paper is as follows. In section 2 we give an introduction to the relevant fusion rules as well as KZ-equations for our study and lay down our methodology for computing braiding matrices. Then, in section 3, we focus on the case of Ising-like Anyons and derive sets of equations governing corresponding conformal blocks. We introduce appropriate coordinates to solve the equations and use connection matrices of hypergeometric functions to transport solutions from one regular singular point to another. This allows us to compute monodromy and subsequently braiding matrices. In section 4 we repeat this process for Fibonacci Anyons and compute corresponding solutions around singular points in configuration space using hypergeometric functions. Again the computation of monodromy and braiding matrices is straightforward once the solutions have been identified in this way. Finally, in section 5 we wrap up by presenting brief conclusions. Appendices contains details of hypergeometric equations relevant for this work as well as some intermediate computational results.

\section{Anyon braiding and KZ equations}

In this section we discuss the modular tensor categories relevant for this work and introduce the techniques we employ to compute braiding matrices.

\subsection{The $\widehat{\mathfrak{su}}(2)_k$ Fusion Categories}

The fusion category $\widehat{\frak{su}}(2)_k$, for $k$ a non-negative integer, consists of $k+1$ so-called simple objects labeled by their spin $0, \half, \ldots, \frac{k}{2}$. Two objects can be \textit{fused} to a third object by morphisms of the category $a \otimes b \rightarrow c$. Then one introduces an integer $N^c_{ab} = \dim \mathrm{mor} (a \otimes b \rightarrow c)$ giving the number of \textit{fusion channels}. One writes
\begin{equation}
    a \otimes b = \bigoplus_c N^c_{ab} c. 
\end{equation}
The simple objects in the category and the fusion coefficients $N^c_{ab}$ have to satisfy several conditions as outlined in \cite{Fusion:2005,Tensor:2015}. In the case of the $\widehat{\frak{su}}(2)_k$-category, the simple objects correspond to chiral conformal primary fields of the corresponding Wess-Zumino-Witten CFT \cite{KNIZHNIK198483,Moore-Seiberg} and the fusion coefficients can be explicitly computed to be
\begin{equation}
    N^{c}_{ab} = \left\{\begin{array}{ll}1 & \quad a + b \geq c,~ b + c \geq a, ~c + a \geq b, ~a + b + c \in \IZ, ~\textrm{ and } a + b + c \leq k\\
    0 & \quad \textrm{otherwise}.\end{array}\right.
\end{equation}
Important instances of this particular fusion category are the Ising category\footnote{Strictly speaking, the $k=2$ category has different R-matrices than the Ising category, but these differ merely in phases and we view the Ising category as the equivalence class of all such categories with identical fusion symbols.} for $k=2$ and the Fibonacci category which arises as a subalgebra of the $k=3$ algebra. Concretely, one has 
\begin{itemize}
    \item \underline{$k=2$}:
    \begin{gather*}
        \mathbf{1} \equiv 0, \quad \sigma \equiv \half, \quad \psi \equiv 1\\
        \sigma \otimes \psi = \psi \otimes \sigma = \sigma, \quad \psi \otimes \psi = \mathbf{1}, \quad \sigma \otimes \sigma = \mathbf{1} \oplus \psi, \quad \mathbf{1} \otimes X = X, \textrm{ for } X=\mathbf{1},~\psi,~\sigma.
    \end{gather*}
    \item \underline{$k=3$}:
    \begin{gather*}
        \mathbf{1} \equiv 0, \quad \tau \equiv 1\\
        \tau \otimes \tau = \mathbf{1} \oplus \tau, \quad \mathbf{1} \otimes X = X \textrm{ for } X=\mathbf{1},~\tau.
    \end{gather*}   
\end{itemize}
We now would like to compute chiral correlation functions of such conformal primaries known as conformal blocks. Taking a more general viewpoint, we can consider conformal blocks $\langle \phi_{\lambda_1}(z_1)\ldots \phi_{\lambda_N}(z_n)\rangle$ of $N$
chiral primaries of highest weight spins $\lambda_1,\ldots, \lambda_N$. Locally, at a point in configuration space, these can be viewed as maps
\begin{equation}
    \psi(z_1,\ldots,z_N) : V_{\lambda_1} \otimes \cdots \otimes V_{\lambda_N} \rightarrow \IC,
\end{equation}
where the $V_{\lambda_i}$ for $i=1,\ldots, N$ are highest weight modules of $\frak{su}(2)$ with highest weight $\lambda_i$. In fact, for given $\lambda_i$, the conformal block is a state in the vector space spanned by \textit{fusion trees}:
\begin{equation} \label{eq:cfblocktree}
    \langle \phi_{\lambda_1}(z_1)\ldots \phi_{\lambda_N}(z_n)\rangle_{\mu_1,\ldots,\mu_{N-3}}=\adjincludegraphics[valign=c]{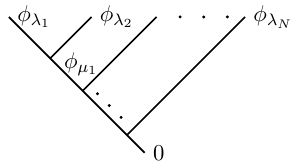}
\end{equation}
More precisely, the Fibonacci category is obtained from the $\mathfrak{su}(2)_3$ category by considering only the integer spin anyons.

These conformal blocks satisfy the Knizhnik-Zamolodchikov equations \cite{KNIZHNIK198483,Etingof:1998ru}
\begin{equation} \label{eq:KZ}
    \kappa \frac{\partial}{\partial z_i}\psi = \left(\sum_{j=1,j\neq i}^N \frac{\Omega_{ij}}{z_j - z_i}\right)\psi, \quad i = 1, \ldots, N,
\end{equation}
where $\psi$ takes values in the space 
\begin{equation}
    V = V_{\lambda_1} \otimes \ldots \otimes V_{\lambda_N},
\end{equation}
and
\begin{equation}
    \kappa = k + h^{\vee}, 
\end{equation}
where $h^{\vee}$ is the dual Coxeter number of the gauge group of the WZW model which in our case is $2$. The tensor $\Omega$ has the form
\begin{equation}
    \Omega = e \otimes f + f \otimes e + \half h \otimes h,
\end{equation}
with $e$, $f$ and $h$ being the Lie algebra elements of $\frak{su}(2)$ satisfying commutation relations
\begin{equation}
    [h,e] = 2e, \quad [h,f] = -2f, \quad [e,f] = h.
\end{equation}

General solutions of KZ-equations \eqref{eq:KZ} have been constructed in \cite{Schechtman:1990zc} in terms abstract integral representations as follows. Let $\mathbf{m}=(m_1,\ldots, m_N)$ be a vector of non-negative integers such that $\sum m_j = m$ and $\mathbf{t} = (t_1,\ldots, t_m)$ complex integration variables. Then the solutions are given as
\begin{equation}
    \Psi_C(\mathbf{z}) = \sum_{\mathbf{m}} \int_C \psi_m(\mathbf{z},\mathbf{t}) \rho_{\mathbf{m}}(\mathbf{z},\mathbf{t}) dt_1 \wedge \cdots \wedge d t_m \cdot e^{m_1}v_1 \otimes \cdots \otimes e^{m_N} v_N,
\end{equation}
where $C$ is a contour in $\mathbf{t}$-space and $v_1, \ldots, v_N$ are lowest weight vectors of $V_{\lambda_1},\ldots, V_{\lambda_N}$. The functions $\psi_m$ and $\rho_{\mathbf{m}}$ are specified by
\begin{eqnarray}
    \rho_{\mathbf{m}}(\mathbf{z},\mathbf{t}) & = & \sum_{s \in \sigma_{\mathbf{m}}} \prod_{n=1}^m (t_n - z_{s(n)})^{-1}, \\
    \psi_m(\mathbf{z},\mathbf{t}) & = & \prod_{i<j} (z_i - z_j)^{\frac{\lambda_i \lambda_j}{2\kappa}} \prod_{p,j}(t_p - z_j)^{-\frac{\lambda_j}{\kappa}} \prod_{p<n} (t_p - t_n)^{\frac{2}{\kappa}},
\end{eqnarray}
where $\sigma_{\mathbf{m}}$ is the set of all maps $s: \{1,\ldots,m\} \rightarrow \{1,\ldots,N\}$ such that the number of elements in $s^{-1}(j)$ equals $m_j$ for all $1 \leq j \leq N$. For generic $\kappa \in \mathbb{C}$ one can show that there are $\binom{N+m-2}{m}$ different contours corresponding to the conformal blocks of $N$-point functions. For $\kappa \in \mathbb{Z}$, which is the case of interest here, many of these contours lead to trivial integrals and the space of conformal blocks reduces to those of the corresponding rational CFT. The solutions we will construct are limiting cases of these integral representations where the particular limit is explained in the following.

\subsection{Braiding from KZ equations}
The KZ-equations can be interpreted as describing a flat connection in a trivial bundle with the fiber $V$ over the configuration space
\begin{equation}
    X_N = \{(z_1,\ldots,z_N) \in \IC | z_i \neq z_j\}.
\end{equation}
In this language, solutions of the KZ equations are flat sections of the corresponding local system. For any given subset $U \subset \IC^N$ we will denote the space of solutions or equivalently the space of flat sections by $\Gamma_f(U,V_{\textrm{KZ}})$. Then, for different open subsets $U_i$ and $U_j$ with $U_i \neq U_j$, the coresponding spaces $\Gamma_f(U_i,V_{\textrm{KZ}})$ and $\Gamma_f(U_j,V_{\textrm{KZ}})$ are related by connection matrices $F_{ij}$.
In order to compute the representation of the braid group as acting on conformal blocks, we need to deduce the monodromy of this local system. To this end, for any path $\gamma: [0,1] \rightarrow X_N$ denote by $M_{\gamma}$ the operator of holonomy along $\gamma$. We are interested in the computation of braid group generators $B_{i,i+1}$. Choosing a base point $\mathbf{z}_0 = (z_1,\ldots,z_N)$ such that $z_i \in \IR,~z_1 > z_2 > \cdots z_N$, the generator $B_{i,i+1}$ describes the exchange of $z_i$ and $z_{i+1}$ along the path depicted below:
\begin{figure}[h]
  \centering
	\includegraphics[width=0.6\textwidth]{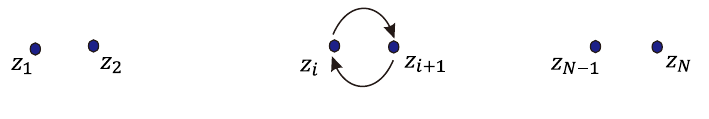}
  \caption{Braiding of two adjacent marked points.}
  \label{fig:braiding}
\end{figure}

\begin{figure}[h]
  \centering
	\includegraphics[width=\textwidth]{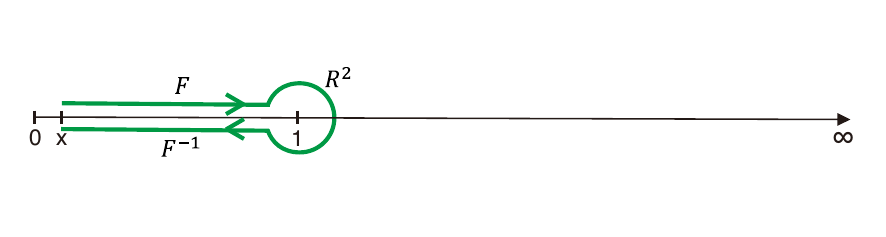}
  \caption{Monodromy operator around $x=1$. We use the $F$-matrix to transport the solutions around $x=0$ to $x=1$, then do a $2\pi i$-rotation around $x=1$ and finally transport the solutions back to $x=0$.}
  \label{fig:monodromy}
\end{figure}

In order to implement braiding, we thus consider the subset $D \subset X_N$ defined by
\begin{equation}
    D = \{(z_1,\ldots, z_N) \in \IR^N | z_1 < z_2 < \cdots < z_N\}, 
\end{equation}
set $z_N = \infty$ and choose coordinates
\begin{equation}
 \label{generalcoordinate}
    \begin{gathered}
           x_1 = \frac{z_2-z_1}{z_3-z_1}, \quad \ldots, \quad x_{N-3} = \frac{z_{N-2}-z_1}{z_{N-1}-z_1}, \\
    x_{N-2} = z_{N-1} - z_1, \quad x_{N-1} = z_1 + \cdots + z_{N-1}.
    \end{gathered}
\end{equation}
Then $(z_1,\ldots,z_{N-1}) \mapsto (x_1,\ldots, x_{N-1})$ is one-to-one and therefore any function $f = f(z)$ on $D$ can be considered as an analytic function of the $x_i$. These coordinates are particularly suitable to describe braiding. To see this, note that taking the limits
\begin{equation} \label{eq:limit}
    x_j \rightarrow 0 \quad \forall \quad j \neq i-1,
\end{equation}
while keeping $x_{i-1}$ finite, one sees that the braiding shown in Fig.  \ref{fig:braiding} can be implemented by letting the coordinate
\begin{equation}
    1 - x_{i-1} = \frac{z_{i+1}-z_i}{z_{i+1}-z_1} \ll 1
\end{equation}
undergo a $\pi i$-rotation. As we will show in the following sections, solutions of KZ-equations under the limit \eqref{eq:limit} can be described in terms of Hypergeometric functions (see Appendix \ref{sec:app} for a detailed discussion). Hypergeometric functions are eigenstates of the monodromy operator and thus, in order to compute the action of the braiding operator, we first compute the monodromy around $1$ as shown in Fig. \ref{fig:monodromy} and then take its square root. 
In order to compute the action of the monodromy operator properly, we need an identification of solutions of KZ-equations to conformal block trees \eqref{eq:cfblocktree}. To this end, we multiply a given solution $\psi$ around the point\footnote{In case of $i-1=1$, we expand solutions around $x_1 \sim 0$.} 
\begin{equation}
    (\ldots,x_{i-1},\ldots) = (0,\ldots 0, 1,0,\ldots,0)
\end{equation}
with $(x_{i-1}-1)^{-(\Delta_{\mu_i}-\Delta_{\lambda_i}-\Delta_{\lambda_{i+1}})}$, where $\Delta_{\lambda}$ is the conformal dimension of the primary $\phi_{\lambda}$, such that
\begin{equation}
    \lim_{x_{i-1}\rightarrow 1} (x_{i-1}-1)^{-(\Delta_{\mu_i}-\Delta_{\lambda_i}-\Delta_{\lambda_{i+1}})} \psi(0,\ldots,0,x_{i-1},0,\ldots,0) = I_{(\lambda_i, \lambda_{i+1},\mu_i)}~,
\end{equation}
where $I_{(\lambda_i,\lambda_{i+1},\mu_i)} \in \textrm{Hom}(V_{\lambda_1} \otimes \cdots \otimes V_{\lambda_N},\IC)$ and can be identified with a particular fusion tree corresponding to representations $\lambda_i$ and $\lambda_{i+1}$ fusing to $\mu_i$. This procedure will be exemplified in more detail as we go along.

\section{Ising Anyons}
Now we apply the technique to a specific model, namely the Ising anyons which we realize here through the $\widehat{\mathfrak{su}}(2)_2$ WZW model. To do this, we need to first briefly review the KZ equation in WZW models.
\subsection{4-point conformal blocks from KZ-equations}
In a WZW model, the correlation functions of fields satisify the KZ equation due to the symmetry of the current algebra. In principle, the explicit form of the correlation functions can be obtained by solving these KZ equations. We start from the analysis of $4$-point functions. Using $\psi$ to denote conformal blocks $\psi(z_i)=\langle \phi_{\lambda_1}(z_1) \phi_{\lambda_2}(z_2)\phi_{\lambda_3}(z_3) \phi_{\lambda_4}(z_4)\rangle$, we have the following KZ equations 
$$\kappa\frac{\partial}{\partial z_i}\psi=\left( \sum_{j=1,j\neq i}^N\frac{\Omega_{ij}}{z_j-z_i}\right)\psi, \quad i=1,...,4\quad .$$
We send $z_4$ to infinity, so $\psi=\psi(z_1,z_2,z_3,\infty)$ satisfies the following:
\begin{align}
	\label{KZ eq}
	\kappa \partial_{z_1}\psi=\left (\frac{\Omega_{12}}{z_2-z_1}+\frac{\Omega_{13}}{z_3-z_1}\right)\psi,\\
	\kappa \partial_{z_2}\psi=\left(\frac{\Omega_{21}}{z_1-z_2}+\frac{\Omega_{23}}{z_3-z_2}\right )\psi,\\
	\kappa \partial_{z_3}\psi=\left (\frac{\Omega_{31}}{z_1-z_3}+\frac{\Omega_{32}}{z_2-z_3}\right)\psi,
\end{align}
where $\Omega_{ij}=\Omega_{ji}$. Now we want to reduce these equations to a simpler form.
To this end, we transform variables as follows:
\begin{equation}
	x=\frac{z_1-z_2}{z_1-z_3},\qquad y=z_1-z_3, \qquad t=z_1+z_2+z_3.
\end{equation}
The inverse transformation is 
\begin{equation}
z_1=\frac13(t+xy+y),\qquad z_2=\frac13(t+xy+y)-xy,\qquad z_3=\frac13(t+xy+y)-y.
\end{equation}
Using these variables, the 3 KZ equations above give us
\begin{equation}
   \kappa \partial_t\psi=0. 
\end{equation}
\begin{equation}
\label{HG to be 1}
-\kappa \partial_x\psi=\left(\frac{\Omega_{21}}{x}+\frac{\Omega_{23}}{x-1}\right)\psi.
\end{equation}
\begin{equation}
\label{HG to be y}
-\kappa\partial_y\psi=\left(\frac{\Omega_{12}+\Omega_{13}+\Omega_{23}}{y}\right)\psi.
\end{equation}
This implies that we can split $\psi$ as follows 
$$\psi(x,y,t)=y^{-(\Sigma \Omega_{ij})/\kappa}F(x),$$
where $F(x)$ satisfies
$$-\kappa \partial_x F=\left(\frac{\Omega_{21}}{x}+\frac{\Omega_{23}}{x-1}\right)F.$$
After choosing a representation for $\Omega_{ij} $, $F(x)$ can be solved from the above equations.
\subsection{Correlators of spin fields}
Ising Anyons can be identified with the $\widehat{\mathfrak{su}}(2)_2$ WZW chiral primaries up to twisting of topological spins as discussed below. For the $\widehat{\mathfrak{su}}(2)_2$ WZW model, the central charge $c=\frac{k \text{dim}\mathfrak{su}(2)}{k+h^{\vee}}=\frac{3}{2}$. The dual coxeter number is  $h^{\vee}=2$ for $\mathfrak{su}(2)$. 
We list here the conformal weights of all the primary fields of the $\widehat{\mathfrak{su}}(2)_2$ WZW model:
\begin{equation}
	\large
	\renewcommand{\arraystretch}{1.2}
\begin{array}{|c|c|c|}	
	\hline
	j(\text{spin})&\lambda&h\\
	\hline
	\frac12&\omega_1&\frac{3}{16}\\
	\hline
	1&2\omega_1&\frac{1}{2}\\
	\hline
\end{array}
\end{equation} 
where $\omega_1$ is the fundamental weight. The conformal weights are computed from $h=\frac{\langle\lambda,\lambda+2\rho\rangle}{2(k+h^{\vee})}$, $\rho=\omega_1$ being the Weyl(principal) vector. The $\sigma$ field in the Ising model is identified with the spin-$\frac12$ field. We want to calculate the correlators of $4$ $\sigma$ fields: $\psi=\langle \mathcal{\sigma}(z_1)\mathcal{\sigma}(z_2)\mathcal{\sigma}(z_3)\mathcal{\sigma}(z_4)\rangle$ in this theory. Each $\mathcal{\sigma}_i$ lives in a fundamental representation of $\mathfrak{su}(2)$, which is the familiar spin-$\frac12$ doublet. We use the symbol $\upa$ and $\doa$ to denote the spin up and down states respectively. $\psi$ lives in the $(\frac{1}{2})\otimes(\frac{1}{2})\otimes(\frac{1}{2})\otimes(\frac{1}{2})$ tensor representation, of which the total dimension is $2^4=16$. The basis of this space is denoted by $\upa\upa\upa\upa,\upa\upa\upa\doa,...$.
This tensor product representation can be decomposed into irreducible representations with definite spins:
$(\frac12)^4=2(0)+3(1)+(2)$, the number in parenthesis being the spin numbers. In order for the correlation function $\psi$ to be invariant under the $SU(2)$ symmetry, the non-zero spin components should be zero. In other words we only need to consider states in the $(0)$ representation. There are 2 copies of $(0)$ in $\psi$, and we need to pick 2 linearly independent basis elements, which is equivalent to choosing a ``fusion channel". For the purpose that will be clear later, here we use the channel specified by the trees below.
\begin{center}
\scalebox{1.2}
{
\begin{tikzpicture}
		
        \draw (0.8,-0.5) node[anchor=west]{$ {I_1 =}$};
		\draw[black, thick] (2,1) -- (4.3,-1.3)node[anchor=west] {$ {0}$};
		\draw (2.1,1) node[anchor=west]{$ {\frac12}$};
		\draw[black, thick] (3.4,1)node[anchor=west]{$ {\frac12}$}--(2.7,0.3);
		\draw[black, thick] (4.5,1)node[anchor=west]{$ {\frac12}$}--(3.25,-0.25);
		\draw (2.8,0.2) node[anchor=west]{$ {0}$};
		\draw[black,thick](5.8,1)node[anchor=west]{$ {\frac12}$}--(3.9,-0.9);

        \draw (7.8,-0.5) node[anchor=west]{$ {I_2 =}$};
		\draw[black, thick] (9,1) -- (11.3,-1.3)node[anchor=west] {$ {0}$};
		\draw (9.1,1) node[anchor=west]{$ {\frac12}$};
		\draw[black, thick] (10.4,1)node[anchor=west]{$ {\frac12}$}--(9.7,0.3);
		\draw[black, thick] (11.5,1)node[anchor=west]{$ {\frac12}$}--(10.25,-0.25);
		\draw (9.8,0.2) node[anchor=west]{$ {1}$};
		\draw[black,thick](12.8,1)node[anchor=west]{$ {\frac12}$}--(10.9,-0.9);
	\end{tikzpicture}
	}
	\end{center}
We refer to \eqref{oldbasis} for their representation as sums of products of  spin $1/2$ states.

Now we can rewrite $\psi$ as:
\begin{equation}
\label{2conformal blocks}
\psi=I_1F_1(z_1,z_2,z_3,z_4)+I_2F_2(z_1,z_2,z_3,z_4),
\end{equation}
where $F$ are scalar functions of $z_i$.
Next, we compute the action of $\Omega_{ij}=(\frac12h\otimes h+f\otimes e+e\otimes f)=2\sum_a t_i^a\otimes t_j^a$ on $I_1$ and $I_2$. This is given in Appendix \ref{4 point expression}.
Setting $\kappa=k+2=4$, we combine \eqref{2conformal blocks} and \eqref{HG to be 1} to get the following equations:
\begin{equation}
\label{KZ equations reduced}
\left(\partial_xF_1-\frac38\frac{F_1}{x}+\frac{\sqrt3}{8}\frac{F_2}{x-1}\right)I_1+\left(\partial_xF_2+\frac{\sqrt 3}{8}\frac{1}{x-1}F_1+\frac14\left(\frac1{2x}-\frac1{x-1}\right)F_2\right)I_2=0,\\
\end{equation}
where $F_1(x),F_2(x)$ come from $F(x)=I_1F_1(x)+I_2F_2(x)$.
$I_1$ and $I_2$ are linearly independent, so the coefficients should vanish simultaneously, that is to say
\begin{gather}
\partial_xF_1(x)=\frac{3 F_1(x)}{8 x}-\frac{\sqrt{3} F_2(x)}{8 (x-1)}, \quad
\partial_x F_2(x)=\frac{1}{4} \left(\frac{1}{x-1}-\frac{1}{2 x}\right) F_2(x)-\frac{\sqrt{3} F_1(x)}{8 (x-1)}.
\end{gather}
To solve this, first we use one equation to express $F_2$ in terms of $F_1$ and subsequently plug it into the other equation, giving
\begin{gather}
F_2(x)=\frac{1-x}{\sqrt3x}(8xF_1'(x)-3F_1(x)),\\
\textrm{and} \nonumber\\
64x(x-1)F_1''(x)+16(1+2x)F_1'(x)-\frac{3(8x-7)}{x(x-1)}F_1(x)=0.
\end{gather} 
Using the local monodromy, we set $F_1(x)=x^{\frac38}(1-x)^{-\frac18}f(x)$ and substitute into the above equation. Then we get a hypergeometric differential equation (see Appendix \ref{sec:app} for details): 
\begin{equation}
x(1-x)f''(x)+(\frac12-x)f'(x)+\frac{1}{16}f(x)=0.
\end{equation}
We can read off the equation parameters $a=\frac14,b=-\frac14,c=\frac12$. Now computing the correlation functions amounts to solving this equation. Fortunately, in this particular case the solutions of the hypergeometric equation are elementary functions.
So we can get $2$ linearly-independent solutions around $x=0$ of the original first-order equation-system:
\begin{gather}
		\left\{
	\begin{aligned}
		F_1^{(+)}&=x^{\frac38}(1-x)^{-\frac18}\frac{\left(\sqrt{1-\sqrt{x}}+\sqrt{\sqrt{x}+1}\right)}{2}\\
		F_2^{(+)}&=\frac{(1-x)\left(\sqrt{\sqrt{x}+1}-\sqrt{1-\sqrt{x}}\right) +\sqrt{x} \left(\sqrt{1-\sqrt{x}}+\sqrt{\sqrt{x}+1}\right)}{\sqrt3 (1-x)^{5/8} x^{1/8}}
	\end{aligned}
	\right.,\\	
	\left\{
	\begin{aligned}
		F_1^{(-)}&=x^{\frac{7}{8}}(1-x)^{\frac{-1}{8}}\frac{\sqrt2}{\sqrt{1+\sqrt{1-x}}}\\
		F_2^{(-)}&=\frac{\sqrt{2} \left(4 \left(\sqrt{1-x}+1\right)-x\left(\sqrt{1-x}+3\right) \right)}{\sqrt3 \left(\sqrt{1-x}+1\right)^{3/2} (x(1-x))^{1/8}}
	\end{aligned}
	\right..
\end{gather}
Computing their asymptotic behavior around $x=0$ gives:
\begin{gather}
		\left\{
		\begin{aligned}
			F_1^{(+)}&\sim x^{3/8}\\
			F_2^{(+)}&\sim \frac{1}{4\sqrt3}x^{11/8}
		\end{aligned}
	\right.,\qquad
		\left\{
		\begin{aligned}
			F_1^{(-)}&\sim x^{7/8}\\
			F_2^{(-)}&\sim \frac{4}{\sqrt3}x^{-1/8}
		\end{aligned}
	\right..
\end{gather}
To relate to the original conformal block $\psi$ viewed as a section of a vector bundle, we write 
\begin{equation}
    F^+=F_1^+I_1+F_2^+I_2,\quad F^-=F_1^-I_1+F_2^-I_2.
\end{equation}
We can see that
\begin{equation}
    \lim_{x\rightarrow 0} x^{-3/8}F^{+} = \left(1\atop 0\right) = I_1, \quad \lim_{x\rightarrow 0}x^{1/8}F^- = \left(0\atop \frac{4}{\sqrt{3}}\right) = \frac{4}{\sqrt{3}} I_2.
\end{equation}
So we can identify $F^+$ with $I_1$ and $F^-$ with $I_2$.
Similarly, we can write down solutions around $x=1$, which will be important in the next section:
\begin{equation}
	\left\{
	\begin{aligned}
		G_1^{(+)}&=\frac{\left(\sqrt{1-\sqrt{1-x}}+\sqrt{\sqrt{1-x}+1}\right) x^{3/8}}{2(1-x)^{1/8}}\\
		G_2^{(+)}&=\frac{(1-\sqrt{1-x})^{\frac32}+ (\sqrt{1-x}+1)^{\frac32}}{2\sqrt3(x(1-x))^{1/8}}
	\end{aligned}
	\right.	,
\end{equation}
\begin{equation}
    	\left\{
	\begin{aligned}
		G_1^{(-)}&=\frac{\sqrt{2} (x(1-x))^{3/8}}{\sqrt{\sqrt{x}+1}}\\
		G_2^{(-)}&=-\frac{\sqrt{2}(2+ \sqrt x) (1-x)^{3/8}}{\sqrt3\sqrt{\sqrt{x}+1} x^{1/8}}~.
	\end{aligned}
	\right.
\end{equation}
Again, we can retain the vector bundle section through
\begin{equation}
    G^+=G^+_1I_1+G^+_2I_2,\quad G^-=G^-_1I_1+G^-_2I_2.
\end{equation}
To examine the scaling behavior of $G^{\pm}$ around $x=1$, we need to transform $(I_1,I_2)$ into another basis, or in other words, we should use a different fusion channel suitable for solutions around $x=1$.  
Here the basis should be chosen to be the eigenbasis of the operator $\Omega_{23}$ corresponding to the following set of trees:
\begin{center}
\scalebox{1.2}
  {
	\begin{tikzpicture}
	
	\draw(-1.9,-0.25)node{$I_1^{'} =$};
	\draw[black, thick] (-1,1)node[anchor=east]{$\frac{1}{2}$} -- (1.3,-1.3)node[anchor=west] {$0$};
	\draw[black, thick] (0,1)node[anchor=east]{$\frac{1}{2}$} --(0.75,0.25);
	\draw[black, thick] (1.5,1)node[anchor=west]{$\frac{1}{2}$}--(0.25,-0.25);
	\draw (0.5,0)node[anchor=west]{$0$} ;
    \draw[black, thick](3,1)node[anchor=west]{$\frac{1}{2}$}--(1,-1);

\draw(5.1,-0.25)node{$I_2^{'} =$};
	\draw[black, thick] (6,1)node[anchor=east]{$\frac{1}{2}$} -- (8.3,-1.3)node[anchor=west] {$0$};
	\draw[black, thick] (7,1)node[anchor=east]{$\frac{1}{2}$} --(7.75,0.25);
	\draw[black, thick] (8.5,1)node[anchor=west]{$\frac{1}{2}$}--(7.25,-0.25);
	\draw (7.5,0)node[anchor=west]{$1$} ;
    \draw[black, thick](10,1)node[anchor=west]{$\frac{1}{2}$}--(8,-1);
	\end{tikzpicture}
	
	}
\end{center}
The exact basis expression and action of the $\Omega$ operators are given in Appendix \ref{4 point expression},
where the old trees and the new ones are related as follows:
\begin{align}
\label{F move basis transformation}
	I_1^{'}=\frac{I_1}{2}-\frac{\sqrt3I_2}{2},\quad I_2^{'}=-\frac{\sqrt3I_1}{2}-\frac{I_2}{2}.
\end{align}
If we use \eqref{F move basis transformation} to change $(I_1,I_2)$ to $(I_1',I_2')$, we get 
\begin{align}
G_1^+I_1+G_2^+I_2&=(\frac{1}{2}G_1^+-\frac{\sqrt3}{2}G_2^+)I'_1+(-\frac{\sqrt3}{2}G_1^+-\frac{1}{2}G_2^+)I'_2\\
&=G^{\prime +}_1I'_1+G^{\prime +}_1I'_2,
\end{align}
where $G^{\prime +}_{1,2}$ have the following scaling behavior:
\begin{equation}
    \left\{
	\begin{aligned}
		G_1^{\prime+}&\sim \frac{15}{4}(1-x)^{7/8}\\
		G_2^{\prime+}&\sim -\frac{2\sqrt3}{3}(1-x)^{-1/8}
	\end{aligned}
	\right..
\end{equation}
So
\begin{equation}
    \lim_{x\rightarrow1}(1-x)^{1/8}G^{+}=-\frac{2\sqrt3}{3}I^{\prime}_2.
\end{equation}
Similarly,
\begin{equation}
    \left\{
	\begin{aligned}
		G_1^{\prime-}&\sim 2(1-x)^{3/8}\\
		G_2^{\prime-}&\sim -\frac{13}{4\sqrt3}(1-x)^{11/8}
	\end{aligned}
	\right.,
\end{equation}
and
\begin{equation}
    \lim_{x\rightarrow1}(1-x)^{-3/8}G^{-}=2I^{\prime}_1.
\end{equation}

\subsection{Conformal blocks and connection matrices}
In the following we introduce some simplifying notations to denote the hypergeometric functions we use (see Appendix \ref{sec:app}):
\begin{align}
    \begin{split}
    u_1&=\prescript{}{2}F_1(a,b,c;z),\\
	u_2&=z^{1-c}\prescript{}{2}{F}_1(a+1-c,b+1-c,2-c;z),\\
	u_3&=\prescript{}{2}{F}_1(a,b,a+b+1-c;1-z),\\
	u_4&=(1-z)^{c-a-b}\prescript{}{2}{F}_1(c-a,c-b,c+1-a-b;1-z).
    \end{split}
\end{align}
They have a linear relation
\begin{align}
	A\begin{pmatrix}
		u_1\\
		u_2
	\end{pmatrix}
=\begin{pmatrix}
    u_3\\
    u_4
\end{pmatrix},
\end{align}
with $A$ a constant matrix.
In our case $a=\frac14,b=-\frac14,c=\frac12$, so according to \eqref{A matrix}
\begin{align}
	A=\left(
	\begin{array}{cc}
		\frac{1}{\sqrt{2}} & \frac{1}{2 \sqrt{2}} \\
		\sqrt{2} & -\frac{1}{\sqrt{2}} \\
	\end{array}
	\right).
\end{align}
We can rewrite the solution we obtained in the last section using hypergeometric functions as follows  
\begin{align}
\label{solution4p}
\begin{split}
\left\{\begin{aligned}
	F_1^{+}&=x^{\frac{3}{8}}(1-x)^{\frac{-1}{8}}u_1\\
	F_2^{+}&=[\frac{1-x}{\sqrt3x}(8x\frac{d}{dx}-3)](x^{\frac{3}{8}}(1-x)^{\frac{-1}{8}})u_1
\end{aligned}\right.
\left\{\begin{aligned}
	F_1^{-}&=x^{\frac{3}{8}}(1-x)^{\frac{-1}{8}}u_2\\
	F_2^{-}&=[\frac{1-x}{\sqrt3x}(8x\frac{d}{dx}-3)](x^{\frac{3}{8}}(1-x)^{\frac{-1}{8}})u_2
\end{aligned}\right.\\
\left\{\begin{aligned}
	G_1^{+}&=x^{\frac{3}{8}}(1-x)^{\frac{-1}{8}}u_3\\
	G_2^{+}&=[\frac{1-x}{\sqrt3x}(8x\frac{d}{dx}-3)](x^{\frac{3}{8}}(1-x)^{\frac{-1}{8}})u_3
\end{aligned}\right.
\left\{\begin{aligned}
	G_1^{-}&=x^{\frac{3}{8}}(1-x)^{\frac{-1}{8}}u_4\\
	G_2^{-}&=[\frac{1-x}{\sqrt3x}(8x\frac{d}{dx}-3)](x^{\frac{3}{8}}(1-x)^{\frac{-1}{8}})u_4
\end{aligned}\right.
\end{split}
\end{align}
Since they are all in the form of linear operators acting on $u_i(x)$'s, and $A$ is a constant matrix, the following relation holds:
\begin{align}
	A\begin{pmatrix}
		F_1^{+}&F_2^{+}\\
		F_1^{-}&F_2^{-}
	\end{pmatrix}
=\begin{pmatrix}
	G_1^{+}&G_2^{+}\\
	G_1^{-}&G_2^{-}
	\end{pmatrix}.
\end{align}

Recall that these solutions are also dependent on the coordinate $y$. Refering back to \eqref{HG to be y}, we see that this dependence can be recovered easily since $I_1$ and $I_2$ are eigenvectors of $\sum\Omega_{ij}$ with the same eigenvalue $-3/2$. $F_i(z_j)$'s and the corresponding $F_i(x)$ differ only by a factor $y^{-3/8}$. From now on we will not distinguish between $\psi(x,y,t)$ and its reduced counterpart $F(x)$.

The conformal blocks of the model are solutions of the KZ equations. They are sections of the vector bundle. $F^{\pm}$ and $G^{\pm}$ in last sections form different basis states of the conformal block space.   
If we rescale them as $F^-\rightarrow\frac12F^{-}$, $G^- \rightarrow\frac12G^{-}$, then by simple linear algebra

\begin{align}
	\begin{pmatrix}
		\mathbf{F}^+&\mathbf{F}^-
	\end{pmatrix}
\frac{1}{\sqrt2}\begin{pmatrix}
	1&1\\1&-1
\end{pmatrix}
=\begin{pmatrix}
	\mathbf{G}^+&\mathbf{G}^-
\end{pmatrix}.
\end{align}
In the following we would call the transfer matrix $\frac{1}{\sqrt2}\begin{pmatrix}
	1&1\\1&-1
\end{pmatrix}$ the ``$F$-move".
\subsection{Braiding matrices}
We have obtained the explicit form of conformal blocks. It turns out they form a representation of the braid group. Roughly speaking, a single braiding move is a ``square root" of a monodromy. Given a complex function $g(z)$, we can multiply the argument with $e^{2\pi i}$, then compare with the original function, which gives the monodromy around $z=0$. Similarly, we can expand $g(z)$ around $z=1$, then multiply $1-z$ with $e^{2\pi i}$ giving the monodromy around $z=1$.

Going back to our conformal blocks, we can see that $F^\pm$ and $G^\pm$ are eigenfunctions of monodromies around $x=0$ and $x=1$ respectively. We can write the action of such a local monodromy as matrices $M_0$ and $M_1$:
\begin{equation}
\begin{pmatrix}
	F^+&F^-
\end{pmatrix}
M_0=
\begin{pmatrix}
	F^+&F^-
\end{pmatrix}
\begin{pmatrix}
	e^{-\frac34\pi i}&0\\
	0&e^{\frac14\pi i}
\end{pmatrix},
\end{equation}
\begin{equation}
\begin{pmatrix}
	G^+&G^-
\end{pmatrix}
M_1=
\begin{pmatrix}
	G^+&G^-
\end{pmatrix}
\begin{pmatrix}
	e^{\frac14\pi i}&0\\
	0&e^{-\frac34\pi i}
\end{pmatrix}.
\end{equation}
Recall that in our new coordinates, $x=\frac{z_1-z_2}{z_1-z_3}$, so that the monodromy $M_0$ is equivalent to  braiding $\sigma(z_1)$ with $\sigma(z_2)$ twice. 

To realize the braiding $B_{12}$ and $B_{23}$ concretely in a matrix form, we need to first define R matrices:
\begin{equation}
	R_{12}=P_{12}\exp\left(\frac{\Omega_{12}\pi i}{4}\right),
\end{equation}
where $P_{12}$ is a permutation.
This operator acts on the local basis states $I_1$ and $I_2$ as:
\begin{equation}
	R_{12}I_1=-\exp\left(-\frac38\pi i\right)I_1,\qquad R_{12}I_2=\exp\left(\frac18\pi i\right)I_2.
\end{equation}
Roughly speaking, acting with $R_{12}$ matrices means swapping the first particle with the second once.

Similarly we can define $R_{23}$:
\begin{align}
    R_{23}I'_1&=P_{23}\exp\left(\frac{\Omega_{23}\pi i}{4}\right)I'_1=-\exp\left(-\frac{3}{8}\pi i\right)I'_1,\\
    R_{23}I'_2&=P_{23}\exp\left(\frac{\Omega_{23}\pi i}{4}\right)I'_2=\exp\left(\frac{1}{8}\pi i\right)I'_2.
\end{align}
Then, in matrix representation,
\begin{equation} \label{eq:R12Ising}
	R_{12}=\begin{pmatrix}
		-e^{-\frac38\pi i}&0\\
		0&e^{\frac18\pi i}
	\end{pmatrix},
\end{equation}
acting from left on the conformal block states $(F^+,F^-)$. We can easily see that $R^2_{12}=M_0$.
Also, we have 
\begin{equation}
	R_{23}=\begin{pmatrix}
		e^{\frac18\pi i}&0\\
		0&-e^{-\frac38\pi i}
	\end{pmatrix},
\end{equation}
acting from left on the basis $(G^+, G^-)$, where again $R^2_{23}=M_1$. To define global braiding matrices $B$, we need to fix a basis, which in our case is $(F^+,F^-)$. So $R_{23}$ needs to be conjugated by the $F$-move to obtain $B_{23}$. Thus we have
\begin{gather}
\label{B12 4}
	B_{12}=R_{12}=\begin{pmatrix}
		-e^{-\frac38\pi i}&0\\
		0&e^{\frac18\pi i}
	\end{pmatrix},\\
B_{23}=FR_{23}F^{-1}=\frac{(-1)^{5/8}}{2}
\label{B23 4}
\begin{pmatrix}
1-i&-1-i\\
-1-i&1-i
\end{pmatrix}.
\end{gather} 
We can check that the braiding relation holds:
\begin{equation}
	B_{12}B_{23}B_{12}=B_{23}B_{12}B_{23}.
\end{equation}
So indeed the conformal blocks form a representation of the braid group. In this case, the braid group has only $2$ generators.

\subsection{6-point conformal blocks}
The $5$-point function of $\sigma$ field vanishes identically in the Ising model. This is because for odd time tensor product of spin $1/2$ representation does not contain spin $0$ component.

So we go straight to $6$-point spin field correlation functions $\psi(z_1...z_6)=\langle \sigma(z_1)...\sigma(z_6)\rangle$.  The goal is also to obtain the braiding matrices of the conformal blocks. Note that, for this purpose the full expression of $6$-point conformal blocks is more than enough. We only need the solution in certain limits where the braided particles are very close and the other particles are far away.
We send $z_6$ to infinity $\infty$, and employ the following change of variables:
$$y=z_5-z_1\quad x=\frac{z_4-z_1}{z_5-z_1}\quad v=\frac{z_3-z_1}{z_4-z_1}\quad w=\frac{z_2-z_1}{z_3-z_1}\quad t=z_1+z_2+z_3+z_4+z_5$$we have $\psi(z_1,z_2,z_3,z_4,z_5,\infty)$. The remaining step is similar to the $4$-point computation so we relegate the details to the Appendix.
The space of conformal blocks is $4$-dimensional, and in the limit of $x,v\rightarrow 0$, the conformal blocks around $w=0$ take the form of
\begin{equation}
    \begin{aligned}
    \mathcal{F}_1= \begin{pmatrix}
F_1^{+}\\
0\\
F_3^{+}\\
0\\
\end{pmatrix},
\mathcal{F}_2=\begin{pmatrix}
F_1^{-}\\
0\\
F_3^{-}\\
0\\
\end{pmatrix},
\mathcal{F}_3=\begin{pmatrix}
0\\
F_2^{+}\\
0\\
F_4^{+}\\
\end{pmatrix},
\mathcal{F}_4=\begin{pmatrix}
0\\
F_2^{-}\\
0\\
F_4^{-}\\
\end{pmatrix}.
\end{aligned}
\end{equation}
The above $4$ vectors are in the basis $\{I_1,I_2,I_3,I_4\}$ of  \eqref{6pbasis}. The $F_{1,2,3,4}^{\pm}$ are all hypergeometric-like functions. In particular, 
\begin{equation}
\begin{aligned}
F_1^+&=F_2^+=w^{3/8}(1-w)^{-1/8}u_1(w)\\
F_3^+&=F_4^+=[\frac{1-w}{\sqrt3w}(8w\frac{d}{dw}-3)](w^{\frac{3}{8}}(1-w)^{\frac{-1}{8}})u_1(w)\\
F_1^-&=F_2^-=w^{3/8}(1-w)^{-1/8}u_2(w)\\
F_3^-&=F_4^-=[\frac{1-w}{\sqrt3w}(8w\frac{d}{dw}-3)](w^{\frac{3}{8}}(1-w)^{\frac{-1}{8}})u_2(w)
\end{aligned}
\end{equation}
Also we have the conformal block around $z=1$:
\begin{equation}
    \begin{aligned}
        \mathcal{G}_1= \begin{pmatrix}
G_1^{+}\\
0\\
G_3^{+}\\
0\\
\end{pmatrix},
\mathcal{G}_2=\begin{pmatrix}
G_1^{-}\\
0\\
G_3^{-}\\
0\\
\end{pmatrix},
\mathcal{G}_3=\begin{pmatrix}
0\\
G_2^{+}\\
0\\
G_4^{+}\\
\end{pmatrix},
\mathcal{G}_4=\begin{pmatrix}
0\\
G_2^{-}\\
0\\
G_4^{-}\\
\end{pmatrix},
    \end{aligned}
\end{equation}
which constitutes the other basis of conformal blocks denoted by $\mathcal{G}_j$, as in the $4$-point case. Here the $G$'s specifically denote:
\begin{equation}
    \begin{aligned}
    G_1^+&=G_2^+=w^{3/8}(1-w)^{-1/8}u_3(w)\\
    G_3^+&=G_4^+=[\frac{1-w}{\sqrt3w}(8w\frac{d}{dw}-3)](w^{\frac{3}{8}}(1-w)^{\frac{-1}{8}})u_3(w)\\
    G_1^-&=G_2^-=w^{3/8}(1-w)^{-1/8}u_4(w)\\
    G_3^-&=G_4^-=[\frac{1-w}{\sqrt3w}(8w\frac{d}{dw}-3)](w^{\frac{3}{8}}(1-w)^{\frac{-1}{8}})u_4(w)
    \end{aligned}
\end{equation}

By rescaling and rearranging the solutions around $w=0$ and $w=1$, we can write the connection matrix between $\mathcal{F}_i$s and $\mathcal{G}_j$s in the following form: 
\begin{equation}
	\frac{1}{\sqrt2}\begin{pmatrix}
	1&1&0&0\\
	1&-1&0&0\\
	0&0&1&1\\
	0&0&1&-1
\end{pmatrix}.
\end{equation} 
The $R$ matrices and braiding matrices are defined similarly to those of the $4$-point case. They both take the form of block diagonal double copies of 4-point functions. The braiding relation
\begin{align}
	B_{12}B_{23}B_{12}=B_{23}B_{12}B_{23}
\end{align}
can easily be checked.

In the 6-point case, the braid group has $4$ generators $B_{i,i+1}$ with $i=1,2,3,4$. To determine $B_{34}$ and $B_{45}$, we need to examine the differential equations corresponding to other two coordinates. This is done explicitly in the appendix.
\begin{figure}
    \centering
     \scalebox{1.2}
{
\begin{tikzpicture}
		
		\draw[black, thick] (2,1) -- (5.5,-2.5);
		\draw (2.1,1) node[anchor=west]{$ {\frac12}$};
		\draw[black, thick] (3.4,1)node[anchor=west]{$ {\frac12}$}--(2.7,0.3)node[anchor=west] {$ \mu_1$};
		\draw[black, thick] (4.5,1)node[anchor=west]{$ {\frac12}$}--(3.25,-0.25)node[anchor=west] {$ \mu_2$};
		\filldraw [black] (5.0,-1.4) circle (0.7pt);
		\filldraw [black] (4.2,-0.6) circle (0.7pt);
		\filldraw [black] (4.6,-1) circle (0.7pt);
		\filldraw [black] (5.7,1) circle (0.7pt);
		\filldraw [black] (6.5,1) circle (0.7pt);
		\filldraw [black] (7.3,1) circle (0.7pt);
		\draw[black, thick] (7.8,1)node[anchor=west]{$ {\frac12}$}--(4.9,-1.9)node[anchor=west] {$ {\mu_{2n-1}=0}$};
	    \draw(8.9,-1.9)node[anchor=west]{$\text{,}$};
	\end{tikzpicture}
	}
    \caption{Fusion tree of $2n$ points}
    \label{fusion tree}
\end{figure}
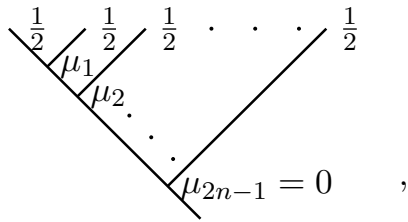
\subsection{General 2n-point conformal blocks}
In general, for $2n$-point function, the corresponding braid group representations $B_{12}$ and $B_{23}$ all take the block diagonal form, built out of $4$-point matrices.

We can explain this fact by generalizing the logic in the $6$-point case. We start from the fusion channel:
\begin{equation}
\label{fusion channel 2n}
	(...((\mathcal O_1\otimes\mathcal O_2)\otimes...\otimes\mathcal{O}_{2n-1})\otimes \mathcal{O}_{2n}),
\end{equation}
This fusion channel can be represented by Figure \ref{fusion tree}. Different conformal blocks correspond to different values of $\mu_i,\; 1\leq i\leq 2n-2$. For each $\mu_i$ there are only $2$ choices: $0$ or $1$. And we fix $\mu_{2n-1}$ to be $0$. So we have $2^{n-2}$ different fusion paths. 

There is no known explicit formula for general $2n$-point spin bases in the tensor product space. Though given a specific $n$, the bases can be computed manually as in the $4$- and $6$-point cases. The essence of the procedure is performing direct sum decomposition of the tensor product repeatedly using the Clebsch-Gordon coefficients. As $n$ increases, this procedure quickly becomes tedious.

But if we only care about the braiding of the first $2$ particles, things simplify a lot. The action of $\Omega_{12}$ and $\Omega_{23}$ only depends on the value of $\mu_1$. If two trees have the same values of $\mu_2,\mu_3...\mu_{2n-2}$, but one of them has $\mu_1=0$, and the other has $\mu_1=1$, the corresponding component function is correlated and decoupled from the others under the coordinate transformation \eqref{generalcoordinate} and the limit $x_{i}\rightarrow 0,\; 2\leq i\leq N-3$. And we have already seen this in the $6$-point case, where $(I_1,I_3)$ $(I_2, I_4)$ form the 2 pairs of basis states and $(F_1,F_3)$ $(F_2,F_4)$ decouple from each other.

From the above argument we can see that general $2n$ conformal blocks look like pairs of hypergeometric-like functions. And there are $2^{n-1}$ of them. Thus the braiding matrices take the form of $2^{n-2}$ copies of blocks \eqref{B12 4} and \eqref{B23 4}. 

\section{$\widehat{\mathfrak{su}}(2)_3$ and Fibonnaci Anyons}
In this section, we will investigate the braid group representation of Fibonacci anyons. Fibonacci anyons play a crucial role in the realization of universal topological quantum computation. The name of this system comes from the fusion rules of the single simple object in the corresponding fusion category, $\tau$: $\tau \otimes \tau = 1 \oplus \tau$, where the quantum dimension of $\tau$ is the golden ratio $\phi=\frac{1+\sqrt{5}}{2}.$ One of the realizations of this anyon is level $3$ $SU(2)$ WZW model, and hence we can apply the techniques developed previously to this case. It should be noted that the Fibonacci category is obtained from $\widehat{\mathfrak{su}}(2)_3$ by considering only the anyons of spin $1$ representations, but the braiding matrices of spin $1$ anyons in these $2$ categories are the same.
\subsection{4-point conformal blocks from KZ equations}
For the $\widehat{\mathfrak{su}}(2)_3$ WZW model, the central charge is $c=\frac{k \text{\ dim}\mathfrak{su}(2)}{k+h^{\vee}}=\frac{9}{5}$.  
We list the conformal weights of all the primary fields here (the notation is the same with the one in the Ising case):
\begin{equation}
	\large
	\renewcommand{\arraystretch}{1.2}
\begin{array}{|c|c|c|}	
	\hline
	j(\text{spin})&\lambda&h\\
	\hline
	\frac12&\omega_1&\frac{3}{20}\\
	\hline
	1&2\omega_1&\frac{2}{5}\\
	\hline
	\frac32&3\omega_1&\frac{3}{4}\\
	\hline
\end{array}
\end{equation} 
The Fibonacci anyon field $\tau$ is identified with the spin $1$ field.
Now we want to repeat the same procedure as in the Ising case to obtain conformal blocks, $F$ and $R$ matrices for Fibonaci anyons. We need to consider the process of fusing $4$ anyons into the vacuum. This is equivalent to distill those spin $0$ states in the $4$-fold tensored spin $1$ representations. The details of the procedure can be found in the Appendix. We get $3$ different spin singles and use $Q$ to denote them:
\begin{equation}
\begin{aligned}
Q_1=(1,0,1,0),\\
Q_2=(1,1,1,0),\\
Q_3=(1,2,1,0), 
\end{aligned}
\end{equation}
where the number in the parenthesis denote total spins after each fusion step. Note here that $Q_3$ is the one with higher intermediate spin than the allowed value imposed by the quantum Clebsch-Gordon rule.

Now we can consider the 4-point KZ-equations, similar with \eqref{KZ eq}. Performing a similar coordinate transformation and imposing the action of $\Omega$, we get the Fuchsian system of differential equations (now $\kappa=k+2=5$):
\begin{equation}
	\label{3 system}
\begin{gathered}
F_1'(x)=\frac{4 F_1(x)}{5 x}-\frac{4 F_2(x)}{5 \sqrt{3} (x-1)},\\
F_2'(x)=-\frac{4 F_1(x)}{5 \sqrt{3} (x-1)}+\frac{1}{5} \left(\frac{2}{x}+\frac{1}{x-1}\right) F_2(x)-\frac{F_3(x)}{\sqrt{15} (x-1)},\\
F_3'(x)=\frac{1}{5} \left(\frac{3}{x-1}-\frac{2}{x}\right) F_3(x)-\frac{F_2(x)}{\sqrt{15} (x-1)}.
\end{gathered}
\end{equation}
At this point we can see that the monodromy matrix at $x=0$ is 
\begin{equation}
	\theta_0=\begin{pmatrix}
		\frac45&0&0\\
		0&\frac25&0\\
		0&0&-\frac25
	\end{pmatrix}.
\end{equation}
Around $x=1$, we have the monodromy matrix (after diagonalisation):
\begin{equation}
	\theta_1=\begin{pmatrix}
		\frac45&0&0\\
		0&\frac25&0\\
		0&0&-\frac25
	\end{pmatrix}.
\end{equation}
$F_2$ and $F_3$ can be solved in terms of $F_1$:
\begin{align}
F_2(x)&=DF_1(x)=-\frac{\sqrt{3} (x-1) \left(5 x F_1'(x)-4 F_1(x)\right)}{4 x},\\
F_3(x)&=D'F_1(x)\nonumber\\
&=\frac{15 (x-1) x \left(5 (x-1) x F_1''(x)-2 (x-3) F_1'(x)\right)+4 (5 (x-6) x+21) F_1(x)}{4 \sqrt{5} x^2},
\end{align}
where $D$ and $D'$ denote the corresponding linear differential operators.
After eliminating $F_2$ and $F_3$ in \eqref{3 system}, we get
\begin{align}
	25 (x-1) x [5 (x-1) x &F_1'''+(7 x+4)F_1'']-5 (3 x (x+14)-46)F_1'\nonumber\\
	&+\frac{4 (x (55 x-107)+56)}{(x-1) x}F_1=0
\end{align}
This is a $3$rd order equation. Using the ansatz $F_1(x)=x^{8/5}(1-x)^{-2/5}f(x)$, we get
\begin{equation}
	\label{hyper?}
x(x-1)f'''(x)+(5 x-4) f''(x)+\frac{99 x^2-154 x+54}{25x(x-1)}f'(x)=0
\end{equation}
$f(x)$ doesn't appear, so in fact this is a $2$nd order equation. Let $H(x)=f'(x)$, and further using an ansatz $H(x)=x^{-9/5}(1-x)^{-1/5}h(x)$, we get a hypergeometric equation:
\begin{equation}
	\label{zhen hyper}
	x(1-x)h''(x)+(\frac25-x)h'(x)+\frac{1}{25}h(x)=0
\end{equation}
Here $a=-\frac15, b=\frac15, c=\frac25$. Around $x=0$, this has 2 solutions:
\begin{equation}
\label{hyperfunction}
\begin{aligned}
u_1(x)&=\prescript{}{2}F_1(-\frac15,\frac15,\frac25;x),\\
u_2(x)&=x^{3/5}\prescript{}{2}F_1(\frac25,\frac45,\frac85;x).
\end{aligned}
\end{equation}
So we get $2$ solutions of the original Fuchsian system:
\begin{equation}
\label{plusminusmeaning}
\begin{gathered}
\mathbf{F}^+=\begin{pmatrix}
F_1^+=x^{8/5}(1-x)^{-2/5}\int_1^xt^{-9/5}(1-t)^{-1/5}u_1(t)dt\\
F_2^+=DF_1^+\\
F_3^+=D'F_1^+
\end{pmatrix},\\
\mathbf{F}^-=\begin{pmatrix}
F_1^-=x^{8/5}(1-x)^{-2/5}\int_1^xt^{-9/5}(1-t)^{-1/5}u_2(t)dt\\
F_2^-=DF_1^-\\
F_3^-=D'F_1^-
\end{pmatrix}.
\end{gathered}
\end{equation}
We notice that each component of the solution is some linear operator acting on the hypergeometric solutions $u_1$ and $u_2$.
The third solution corresponds to $H(x)=const.$, which gives
\begin{gather}
\mathbf{F}^0=
\begin{pmatrix}
	F_1^0=x^{8/5}(1-x)^{-2/5}\\
	F_2^0=DF_1^0\\
	F_3^0=D'F_1^0
\end{pmatrix}	
\end{gather}

The scaling behavior around $x=0$ of $\mathbf{F}^{\pm,0}$ is as follows:
\begin{gather}
	\left\{
	\begin{aligned}
		F_1^{+}&\sim x^{4/5}\\
		F_2^{+}&\sim x^{9/5}\\
		F_3^{+}&\sim x^{14/5}
	\end{aligned}
	\right.\qquad
	\left\{
	\begin{aligned}
		F_1^{-}&\sim x^{7/5}\\
		F_2^{-}&\sim x^{2/5}\\
		F_3^{-}&\sim x^{7/5}
	\end{aligned}
	\right.\qquad
	\left\{ \begin{aligned}
		F_1^{0}&\sim x^{8/5}\\
		F_2^{0}&\sim x^{3/5}\\
		F_3^{0}&\sim x^{-2/5}
	\end{aligned}\right.
\end{gather}
$\mathbf{F}_0$ should be discarded due to its association with $Q_3$. So the conformal block space is two-dimensional. Now we can compute the braiding matrices. We need the action of $R_{12}=P_{12}\exp(\pi i\Omega_{12}/5)$ on $Q_1$ and $Q_2$:
\begin{equation}
R_{12}Q_1=\exp(-\frac45\pi i)Q_1,\quad R_{12}Q_2=-\exp(-\frac25\pi i)Q_2.
\end{equation}
Then we have
\begin{equation}
\label{11block}
    B_{12}=R_{12}=\begin{pmatrix}
	\exp(-\frac45 \pi i)&0\\
	0&-\exp(-\frac{2}{5}\pi i)
\end{pmatrix}
=\begin{pmatrix}
	    q^2&0\\
	    0&-q
	\end{pmatrix},
\end{equation}
where $q=\exp(-2/5 \pi i)$.
From \eqref{zhen hyper} we know that $\mathbf{G}^{\pm}$ around $x=1$ takes the same form as $\mathbf{F}^{\pm}$, with the replacements $u_1\rightarrow u_3, u_2\rightarrow u_4$, thus:
\begin{align}
	u_3(x)&=\prescript{}{2}F_1(-\frac15,\frac15,\frac35;1-x),\\
	u_4(x)&=(1-x)^{2/5}\prescript{}{2}F_1(\frac35,\frac15,\frac75;1-x),
\end{align}
\begin{gather}
\mathbf{G}^+=\begin{pmatrix}
G_1^+=x^{8/5}(1-x)^{-2/5}\int_1^xt^{-9/5}(1-t)^{-1/5}u_2(t)dt\\
G_2^+=DG_1^+\\
G_3^+=D'G_1^+
\end{pmatrix},\\
\mathbf{G}^-=\begin{pmatrix}
G_1^-=x^{8/5}(1-x)^{-2/5}\int_1^xt^{-9/5}(1-t)^{-1/5}u_4(t)dt\\
G_2^-=DG_1^-\\
G_3^-=D'G_1^-
\end{pmatrix}.
\end{gather}
Around $x=1$,
\begin{equation}
\left\{
\begin{aligned}
G^{+}_1&\sim -\frac54(1-x)^{2/5}\\
G^{+}_2&\sim \frac{5\sqrt3}{8}(1-x)^{2/5}\\
G^{+}_3&\sim \frac{5\sqrt5}{8}(1-x)^{2/5}
\end{aligned}
\right.
,\quad 
\left\{
\begin{aligned}
 G^{-}_1&\sim -\frac56(1-x)^{4/5}\\
 G^{-}_2&\sim \frac{5}{2\sqrt3}(1-x)^{4/5}\\
 G^{-}_3&\sim -\frac{5\sqrt5}{6}(1-x)^{4/5}
\end{aligned}
\right.
\end{equation}
We can do a basis transformation $Q\rightarrow Q'$:
\begin{gather}
    Q'_1=\frac13Q_1-\frac{1}{\sqrt3}Q_2+\frac{\sqrt5}{3}Q_3,\\
    Q'_2=-\frac{1}{\sqrt3}Q_1+\frac12Q_2+\frac{1}{2}\sqrt{\frac53}Q_3,\\
    Q'_3=\frac{\sqrt5}{3}Q_1+\frac12\sqrt{\frac{5}{3}}Q_2+\frac16Q_3,
\end{gather}
with $P_{23}Q'_1=Q'_1$, $P_{23}Q'_2=-Q'_2$. The action of $\Omega$ on these new state is in the appendix \eqref{tableQprime}.
So
\begin{equation}
	\label{R matrix anomaly}
	R_{23}=\begin{pmatrix}
		-\exp(-\frac25\pi i)&0\\
		0&\exp(-\frac45\pi i)
	\end{pmatrix}.
\end{equation}
The linear relation between $\mathbf{F}^{\pm}$ and $\mathbf{G}^{\pm}$ can be written as
\begin{equation}
	\begin{pmatrix}
		\mathbf{F}^+&\mathbf{F}^-
	\end{pmatrix}
	A^T
	=\begin{pmatrix}
		\mathbf{G}^+&\mathbf{G}^-
	\end{pmatrix},
\end{equation}  
with the transition matrix
\begin{equation}
	A=\left(
	\begin{array}{cc}
		\frac{\Gamma \left(\frac{3}{5}\right)^2}{\Gamma \left(\frac{2}{5}\right) \Gamma \left(\frac{4}{5}\right)} & \frac{1}{6} \left(\sqrt{5}-1\right) \\
		\sqrt{5}-1 & \frac{\Gamma \left(-\frac{3}{5}\right) \Gamma \left(\frac{7}{5}\right)}{\Gamma \left(\frac{1}{5}\right) \Gamma \left(\frac{3}{5}\right)} \\
	\end{array}
	\right).
\end{equation}
We have the gamma function relations:
$\frac{\Gamma \left(\frac{3}{5}\right)^2}{\Gamma \left(\frac{2}{5}\right) \Gamma \left(\frac{4}{5}\right)}\frac{\Gamma \left(-\frac{3}{5}\right) \Gamma \left(\frac{7}{5}\right)}{\Gamma \left(\frac{1}{5}\right) \Gamma \left(\frac{3}{5}\right)}=\frac13(1-\sqrt5)=-\frac23\phi^{-1}$, where $\phi=\frac{1+\sqrt5}{2}$ denotes the golden ratio.
Let $m=\sqrt\phi A_{11}=\phi^{1/2}\Gamma(\frac35)^2/\Gamma(\frac25)\Gamma(\frac45)$, then by rescaling and rearranging the solution: $$\mathbf{F}^+\rightarrow m\mathbf{F}^+ ,\quad \mathbf{F}^-\rightarrow \frac13\mathbf{F}^- \quad\mathbf{G}^-\rightarrow-m\mathbf{G}^-,$$ we can transform $A^{T}$ to
\begin{equation}
	A^{T}\sim\mathcal{F}=\begin{pmatrix}
		\phi^{-\frac12}&\phi^{-1}\\
		-\phi^{-1}&\phi^{-\frac12}
	\end{pmatrix}.\\
\end{equation} 
The braiding matrices $B_{23}$ is 
\begin{align}
	B_{23}&=\mathcal{F}R_{23}\mathcal{F}^{-1}\\
	\label{22block}
	&=\frac{q}{q+1}\begin{pmatrix}
		-1&q\sqrt{\phi}\\
		q\sqrt{\phi}&q^2
	\end{pmatrix},
\end{align}
where we have used the fact that $\phi=q+q^{-1}+1$.
We also have the braiding relation:
\begin{equation}
B_{12}B_{23}B_{12}-B_{23}B_{12}B_{23}=0
\end{equation}
These results agree with those in  \textit{Nayak et al.}\cite{Nayak_2008}, where $R$ matrix and $F$ matrix of Fibonacci anyons are presented. The pentagon and hexagon relation is verified in detail in \cite{Trebst_2008}.  
\subsection{Braiding matrices for 6-point conformal blocks}
We will briefly list the results here. The spirit is all the same with previous sections.
There are a total of $15$ basis states in this case. The explicit form of these trees in terms of the tensor product is complicated so we will not write out details here. But we can list their fusion path label:
\begin{gather}
\label{15 basis}
\begin{split}
	Q_1=(1,0,1,0,1,0);\quad Q_2=(1,0,1,1,1,0);\quad Q_3=(1,1,0,1,1,0);\\
	Q_4=(1,1,1,0,1,0);\quad Q_5=(1,1,1,1,1,0);\quad Q_6=(1,0,1,2,1,0);\\
	Q_7=(1,2,1,0,1,0);\quad Q_8=(1,2,1,1,1,0);\quad Q_9=(1,1,1,2,1,0);\\
	Q_{10}=(1,1,2,1,1,0);\quad Q_{11}=(1,2,1,2,1,0);\quad Q_{12}=(1,2,2,1,1,0);\\
	Q_{13}=(1,1,2,2,1,0);\quad Q_{14}=(1,2,2,2,1,0);\quad Q_{15}=(1,2,3,2,1,0).
\end{split}
\end{gather}
Thus the correlation function can be written as $\psi=\sum_{1}^{15}F_i(w,v,x)Q_i$. If we send $v,x\rightarrow 0$, there are in total $15$ reduced KZ equations. We can divide them into $7$ decoupled groups: $(F_1,F_4,F_7),(F_2,F_5,F_8),(F_6,F_9,F_{11})$; $(F_{10},F_{12}),(F_{13},F_{14})$; $(F_3),(F_{15})$. Groups with equal number of elements satisfy the same equation. They are as follows:
\begin{equation}
\label{3copluedeq}
\begin{gathered}
	F_1'(w)=\frac{4 F_1(w)}{5 w}-\frac{4 F_4(w)}{5 \sqrt{3} (w-1)},\\
	F_4'(w)=-\frac{4 F_1(w)}{5 \sqrt{3} (w-1)}+\frac{1}{5} \left(\frac{2}{w}+\frac{1}{w-1}\right) F_4(w)-\frac{F_7(w)}{\sqrt{15} (w-1)},\\
	F_7'(w)=\frac{1}{5} \left(\frac{3}{w-1}-\frac{2}{w}\right) F_7(w)-\frac{F_4(w)}{\sqrt{15} (w-1)},
\end{gathered}
\end{equation}
\begin{equation}
\begin{gathered}
F_{10}'(w)=\frac{1}{5}(-\frac{1}{w-1}+\frac{2}{w})F_{10}-\frac{\sqrt3}{5}\frac{F_{12}}{w-1},\\
F_{12}'(w)=\frac{1}{5}(\frac{1}{w-1}-\frac{2}{w})F_{12}-\frac{\sqrt3}{5}\frac{F_{10}}{w-1},
\end{gathered}
\end{equation}
\begin{gather}
F_3'(w)=\frac15(\frac{2}{w-1}+\frac{2}{w})F_3.
\end{gather}
As can be seen from the fusion path of the basis, $Q_i$ with $i>5$ are not allowed because there exist states with intermediate spin higher than $1$. So after solving the equations, we should restrict ourselves to solutions associated with $Q_i,\;i\leq5$. Now the conformal block space is spanned by the following $5$ basis states:
\begin{equation}
\begin{pmatrix}
F_1^+\\
0\\
0\\
F_4^+\\
0\\
0\\
F_7^+\\
0\\
(7\;zeros)
\end{pmatrix},
\begin{pmatrix}
F_1^-\\
0\\
0\\
F_4^-\\
0\\
0\\
F_7^-\\
0\\
...
\end{pmatrix},
\begin{pmatrix}
0\\
F_2^+\\
0\\
0\\
F_5^+\\
0\\
0\\
F_8^+\\
...
\end{pmatrix},
\begin{pmatrix}
0\\
F_2^-\\
0\\
0\\
F_5^-\\
0\\
0\\
F_8^-\\
...
\end{pmatrix},
\begin{pmatrix}
0\\
0\\
F_3\\
0\\
0\\
0\\
0\\
0\\
...
\end{pmatrix},
\end{equation}
with 
\begin{equation}
    \begin{gathered}
        F_1^+=F_2^+=w^{8/5}(1-w)^{-2/5}\int_1^wt^{-9/5}(1-t)^{-1/5}u_1(t)dt,\\
        F_1^-=F_2^-=w^{8/5}(1-w)^{-2/5}\int_1^wt^{-9/5}(1-t)^{-1/5}u_2(t)dt,\\
        F_4^{\pm}=F_5^{\pm}=DF_1^{\pm},\\
        F_7^{\pm}=F_8^{\pm}=D'F_1^{\pm},\\
        F_3=w^{2/5}(w-1)^{2/5},\\
        D=\frac{\sqrt3}{4}(1-w)(5\frac{d}{dw}+\frac{4}{w}),\\
        D'=\frac{15\sqrt5}{4}(w-1)^2\frac{d^2}{dw^2}-\frac{3\sqrt5}{2w}(w-1)(w-3)\frac{d}{dw}+\frac{5w(w-6)+21}{\sqrt 5 w^2}.
    \end{gathered}
\end{equation}
In this basis, the first two braiding matrices are given by 
\begin{equation}
\label{braiding 6 anyon}
\begin{aligned}
	B_{12}&=q\begin{pmatrix}
		q&0&0&0&0\\
		0&-1&0&0&0\\
		0&0&q&0&0\\
		0&0&0&-1&0\\
		0&0&0&0&-1
	\end{pmatrix},
	\\
B_{23}&=\frac{q}{q+1}\begin{pmatrix}
	-1&q\sqrt{\phi}&0&0&0\\
	q\sqrt{\phi}&q^2&0&0&0\\
	0&0&-1&q\sqrt{\phi}&0\\
	0&0&q\sqrt{\phi}&q^2&0\\
	0&0&0&0&-(1+q)
\end{pmatrix}.
\end{aligned}
\end{equation}

\subsection{Braiding matrices of general $m$-point function of Fibonacci anyons}
The dimension of conformal block spaces of $m$-point Fibonacci anyons for arbitrary positive integer $m$ is the $(m-1)$-th Fibonacci number $f_{m-1}$ \cite{Nayak:2008zza} ($f_1=f_2=1,f_3=2,f_3=3...$). In this general case, we can also write the braiding matrices $B_{12}$ and $B_{23}$ in block diagonal form as in the 6-point case just by choosing appropriate basis states.

For general $m$ point fusion trees, let the fusion path be labelled by $(1,\mu_1,\mu_2,...,\mu_{m-1}=0)$. Each $\mu_i$ may take 3 values $0,1,2$. And each tree that satisfies the fusion rule and quantum Clesbch-Gordon rule corresponds to one conformal block. Now the action of $B_{12}$ and $B_{23}$ only depends on the value of $\mu_1$ and $\mu_2$.  

There are trees that always appear as a triplet:
\begin{equation}
\label{treetrio}
    \begin{aligned}
    (1,0,1,...,\mu_i,...,0)\\
    (1,1,1,...,\mu_i,...,0)\\
    (1,2,1,...,\mu_i,...,0)\\
    \end{aligned}
\end{equation}
with same $\mu_i$ in all $3$ lines.
One such triplet will contribute one $2\times 2$ block in \eqref{11block} and \eqref{22block} to the combined braiding matrix. There are $f_{m-3}$ such blocks because deleting the first two path labels in each line of \eqref{treetrio} will create new allowed paths with two particles less.

Besides the above triplets, also paths of the form $(1,1,0,1,...,0)$ will appear. They can't form a triplet because labels $(1,0,0,1,...0)$ and $(1,2,0,1,...0)$ violate spin addition rules. The block corresponding to this path will contribute a single phase $-q$ to the braiding matrix. And there are $f_{m-4}$ of them because deleting the first $3$ labels would create an allowed path in a $(m-3)$-tree.

We can verify that other paths will not contribute to the conformal block space. In conclusion, for general $m$-point Fibonacci anyon correlation functions, the braiding matrices are of rank $f_{m-1}$ with $f_{m-3}$  $2\times 2$ blocks and $f_{m-4}$ single elements $-q$, see \eqref{braiding 6 anyon}. For example, in the $5$-point case,
\begin{equation}
    \begin{aligned}
       B_{12}&=q\begin{pmatrix}
            -1&0&0\\
            0&q&0\\
            0&0&-1
        \end{pmatrix},
        \\ B_{23}&=\frac{q}{q+1}\begin{pmatrix}
          -1&q\sqrt\phi&0\\
          q\sqrt\phi&q^2&0\\
          0&0&-(1+q)
         \end{pmatrix}.
    \end{aligned}
\end{equation}
\section{Conclusions}
In this paper we solved the KZ-equations for non-Abelian $\mathfrak{su}(2)$ level $2$ and level $3$ Anyons around regular singular points in moduli space. We explicitly showed that the monodromy representation captures the quantum statistics of these Anyons in a consistent way. Our analysis only involved 4-point and 6-point functions as the braid group representations of higher point functions can be built up from these. Nevertheless it would be interesting, if not important, to construct explicit solutions valid all over moduli space for any $2n$-point conformal block. This is easier to achieve for the Ising model and corresponding conformal blocks have been constructed in reference \cite{Ardonne:2010hj}. The expressions the authors derive there are purely algebraic in terms of positions of chiral operators which allows to write down an explicit algorithm to compute any such conformal block. The main reason this is possible in the case of the Ising model is that the corresponding hypergeometric function there has algebraic solutions. Such solutions only arise for very particular choices of parameters $a$, $b$ and $c$ (see \cite{Matsuda1985LecturesOA} for a thorough analysis of the matter) which is what happens in the case of the Ising model. In the Fibonacci case, the situation is more complicated as already for 4-point conformal blocks the parameters $a$, $b$ and $c$ do not fall into the classification scheme for algebraic solutions. To make headway for explicit solutions valid everywhere for higher point blocks, it is thus necessary to device new computational mechanisms. Apart from possible analytic methods, it would be interesting here to also consider novel machine learning techniques to solve for such functions. One such approach would be to employ a Restricted Boltzmann Machine (RBM) to solve for groundstates of a Hamiltonian system realizing the topological phase in question \cite{Noormandipour:2020dqp}. In our case this would mean viewing the KZ-connection itself as a Hamiltonian and solving for its horizontal sections by putting the equation on the lattice and performing gradient descent which we leave as project for the future. 

\subsection*{Acknowledgments}
We would like to thank Jingyuan Chen, Emanuel Scheidegger, Youran Sun and Qingrui Wang for valuable discussions. This work is supported by the National Thousand-Young-Talents Program of China. 

\appendix
\section{Hypergeometric equations} \label{sec:app}
The hypergeometric equation (see \cite{Beukers1989} for a nice introduction to the subject) is the second order linear differential equation on the extended complex plane $\mathbb{C}\sqcup\{\infty\}$ of the form
\begin{equation}
    [z(1-z)\partial_z^2+(c-(a+b+1)z)\partial_z-ab]f(z)=0
\end{equation}
where the numbers $a,b,c$ are called the parameters of the hypergeometric equation. 

The 2 linearly independent solution of this equation around the point $z=0$ are
\begin{align}
    \begin{split}
    u_1&=\prescript{}{2}F_1(a,b,c;z)\\
	u_2&=z^{1-c}\prescript{}{2}{F}_1(a+1-c,b+1-c,2-c;z)\\
    \end{split}
\end{align}
where $\prescript{}{2}F_1(a,b,c;z)$ are Euler-Gauss hypergeometric functions defined as
\begin{equation}
   \prescript{}{2}F_1(a,b,c;z)=F(a,b,c,z)=\sum_{m=0}^{\infty}\frac{(a)_m(b)_m}{(c)_m m!}z^m
\end{equation}
with 
\begin{equation}
(\alpha)_m=\alpha(\alpha+1)...(\alpha+m-1)=\frac{\Gamma(\alpha+m)}{\Gamma(\alpha)}    
\end{equation}
the Pochhammer symbol.

The solutions around the point $z=1$ are
\begin{align}
    \begin{split}
   u_3&=\prescript{}{2}{F}_1(a,b,a+b+1-c;1-z)\\
	u_4&=(1-z)^{c-a-b}\prescript{}{2}{F}_1(c-a,c-b,c+1-a-b;1-z)
    \end{split}
\end{align}

Solutions around 2 points are in the same solution space of a equation, in other words, they are different bases of the same $2$ dimensional vector space. One can see that by simply substituting these two set of solutions or use the Riemann scheme, which are characterizing properties of solutions of hypergeometric equations. By properties of differential equations these $2$ set of solutions have a linear relation:
\begin{equation}
	A\begin{pmatrix}
		u_1\\
		u_2
	\end{pmatrix}
=\begin{pmatrix}
    u_3\\
    u_4
\end{pmatrix}
\end{equation}

where $A$ is a constant matrix. We can derive all the elements of $A$  using the Euler integral representation, which is:
\begin{equation}
    F\left( a,b,c,z\right)=\frac{\Gamma (c) }{\Gamma (b) \Gamma (c-b)}\int_0^1 t^{b-1} (1-t z)^{-a} (1-t)^{-b+c-1} \, dt
\end{equation}

If we substitute the linear equation of these $2$ basis into the equation, multiplying the factors of $(1-z)$ on the hypergeometric functions on both side and take the limit $z\rightarrow1$ , then by properties of gamma functions and hypergeometric functions :
\begin{equation}
    \frac{\Gamma (a) \Gamma (b)}{\Gamma (a+b)}=\int_0^1 t^{a-1} (1-t)^{b-1} \, dt ,\quad F(a,b,c,1)=\frac{\Gamma (c) \Gamma (-a-b+c)}{\Gamma (c-a) \Gamma (c-b)}
\end{equation}

We obtain
\begin{equation}
\label{A matrix}
     A=\begin{pmatrix}
\frac{\Gamma(a+b+1-c)\Gamma(1-c)}{\Gamma(a+1-c)\Gamma(b+1-c)}&\frac{\Gamma(a+b+1-c)\Gamma(c-1)}{\Gamma(a)\Gamma(b)}\\
\frac{\Gamma(c+1-a-b)\Gamma(1-c)}{\Gamma(1-a)\Gamma(1-b)}&\frac{\Gamma(c+1-a-b)\Gamma(c-	1)}{\Gamma(c-a)\Gamma(c-b)}
\end{pmatrix}
\end{equation}
\section{Tensor product basis of 4 point $
\mathfrak{su}(2)_2$ conformal blocks}\label{4 point expression}
In section 3.2, the basis that is firstly mentioned is given as follows, the ups and downs illustrate the z-axis angular momentum.
\begin{equation}
\label{oldbasis}
    \begin{gathered}
     I_1=\frac12(\upa\doa\upa\doa+\doa\upa\doa\upa-\doa\upa\upa\doa-\upa\doa\doa\upa),\\
I_2=\frac{1}{\sqrt3}(\upa\upa\doa\doa+\doa\doa\upa\upa)-\frac{1}{2\sqrt3}(\upa\doa\upa\doa+\doa\upa\doa\upa+\doa\upa\upa\doa+\upa\doa\doa\upa).
    \end{gathered}
\end{equation}

Basically, these are obtained by repeated computation with Clebsch-Gordon coefficients. At each step of tensoring, the rules of addition of angular momentum should be obeyed. We can label the 2 basis states by the total angular momentum of the intermediate steps. For example, $I_1=(\frac12,0,\frac12,0)$, $I_2=(\frac12,1,\frac12,0)$. Pictorially, they are
\begin{center}
\scalebox{1.2}
{
\begin{tikzpicture}
		
        \draw (0.8,-0.5) node[anchor=west]{$ {I_1 =}$};
		\draw[black, thick] (2,1) -- (4.3,-1.3)node[anchor=west] {$ {0}$};
		\draw (2.1,1) node[anchor=west]{$ {\frac12}$};
		\draw[black, thick] (3.4,1)node[anchor=west]{$ {\frac12}$}--(2.7,0.3);
		\draw[black, thick] (4.5,1)node[anchor=west]{$ {\frac12}$}--(3.25,-0.25);
		\draw (2.8,0.2) node[anchor=west]{$ {0}$};
		\draw[black,thick](5.8,1)node[anchor=west]{$ {\frac12}$}--(3.9,-0.9);

        \draw (7.8,-0.5) node[anchor=west]{$ {I_2 =}$};
		\draw[black, thick] (9,1) -- (11.3,-1.3)node[anchor=west] {$ {0}$};
		\draw (9.1,1) node[anchor=west]{$ {\frac12}$};
		\draw[black, thick] (10.4,1)node[anchor=west]{$ {\frac12}$}--(9.7,0.3);
		\draw[black, thick] (11.5,1)node[anchor=west]{$ {\frac12}$}--(10.25,-0.25);
		\draw (9.8,0.2) node[anchor=west]{$ {1}$};
		\draw[black,thick](12.8,1)node[anchor=west]{$ {\frac12}$}--(10.9,-0.9);
	\end{tikzpicture}
	}
	\end{center}
We need to compute explicitly the action of $\Omega_{ij}=(\frac12h\otimes h+f\otimes e+e\otimes f)=2\sum_a t_i^a\otimes t_j^a$ on $I_1$ and $I_2$. In the case of $\mathfrak{su}(2)$, we can choose $t^a$ to be $J_z,J_x,J_y$, the familiar angular momentum operators obeying the $\frak{su}(2)$-algebra $[J_a,J_b] = 2 i \epsilon_{abc} J_c$. We have $J_z\uparrow=\frac12\upa,J_x\uparrow=\frac12\doa,J_y\uparrow=\frac i2\doa$, $J_z\doa=-\frac12\doa,J_x\doa=\frac12\upa,J_y\doa=-\frac i2\upa$.
So
$$\Omega_{ij}\upa_i\upa_j=2(J_z\upa_i J_z\upa_j+J_x\upa_i J_x\upa_j+J_y\upa_i J_y\upa_j)=\frac{1}{2}\upa_i\upa_j,$$
$$\Omega_{ij}\upa_i\doa_j=-\frac{1}{2}\upa_i\doa_j+\doa_i\upa_j,$$
$$\Omega_{ij}\doa_i\upa_j=-\frac{1}{2}\doa_i\upa_j+\upa_i\doa_j,$$
$$\Omega_{ij}\doa_i\doa_j=\frac{1}{2}\doa_i\doa_j.$$
\begin{equation}
		\large
	\renewcommand{\arraystretch}{1.2}
\begin{array}{c|c|c}
 \text{} & I_1 & I_2 \\ \hline
 \Omega _{12} & -\frac{1}{2} \left(3 I_1\right) & \frac{I_2}{2} \\ \hline
 \Omega _{13} & -\frac{1}{2} \sqrt{3} I_2 & -\frac{1}{2} \sqrt{3} I_1-I_2 \\ \hline
 \Omega _{14} & \frac{\sqrt{3} I_2}{2} & \frac{\sqrt{3} I_1}{2}-I_2 \\\hline
 \Omega _{23} & \frac{\sqrt{3} I_2}{2} & \frac{\sqrt{3} I_1}{2}-I_2 \\\hline
 \Omega _{24} & -\frac{1}{2} \sqrt{3} I_2 & -\frac{1}{2} \sqrt{3} I_1-I_2 \\\hline
 \Omega _{34} & -\frac{1}{2} \left(3 I_1\right) & \frac{I_2}{2} \\\hline
\end{array}
\end{equation}

For the new basis, we have:
\begin{equation}
\label{newbasis}
    \begin{gathered}
    I_1^{'}=\frac12(\upa\doa\upa\doa+\doa\upa\doa\upa-\upa\upa\doa\doa-\doa\doa\upa\upa),\\
I_2^{'}=\frac{1}{\sqrt3}(\upa\doa\doa\upa+\doa\upa\upa\doa)-\frac{1}{2\sqrt3}(\upa\doa\upa\doa+\doa\upa\doa\upa+\upa\upa\doa\doa+\doa\doa\upa\upa),
    \end{gathered}
\end{equation}
and the action of $\Omega_{ij}$'s on the new basis states is:
\begin{equation}
		\large
\renewcommand{\arraystretch}{1.2}
	\begin{array}{c|c|c}
		\text{} & I'_1 & I'_2 \\\hline
		\Omega _{12} & \frac{\sqrt{3} I'_2}{2} & \frac{\sqrt{3} I'_1}{2}-I'_2 \\\hline
		\Omega _{13} & -\frac{1}{2} \sqrt{3} I'_2 & -\frac{1}{2} \sqrt{3} I'_1-I'_2 \\\hline
		\Omega _{14} & -\frac{1}{2} \left(3 I'_1\right) & \frac{I'_2}{2} \\\hline
		\Omega _{23} & -\frac{1}{2} \left(3 I'_1\right) & \frac{I'_2}{2} \\\hline
		\Omega _{24} & -\frac{1}{2} \sqrt{3} I'_2 & -\frac{1}{2} \sqrt{3} I'_1-I'_2 \\\hline
		\Omega _{34} & \frac{\sqrt{3} I'_2}{2} & \frac{\sqrt{3} I'_1}{2}-I'_2 \\\hline
	\end{array}
\end{equation}
\section{Comparison to standard Ising braiding matrices}
Here we want to compare the braiding matrices obtained above to those of standard Ising anyons. Ising braiding matrices have a simple representation using Majorana Fermions\cite{Ivanov:2001}.
Let $\gamma_i(1\leq i\leq 2N)$ satisfy the anticommutation relation $\{\gamma_i,\gamma_j\}=2\delta_{ij}$. First, we consider the simplest case $N=1$. Let $\Psi=(\gamma_1+i\gamma_2)/2, \Psi^{\dagger}=(\gamma_1-i\gamma_2)/2$. It can be verified that $\Psi^2=(\Psi^{\dagger})^2=0$.  We define the vacuum $\ket{0}$ as $\Psi\ket{0}=0$. Choosing $\ket{0}$ and $\Psi^{\dagger}\ket{0}$ as the basis, we can represent the operators as follows:
\begin{gather}
	\Psi\doteq\begin{pmatrix}
		0&1\\0&0
	\end{pmatrix}
,\Psi^{\dagger}\doteq\begin{pmatrix}
	0&0\\1&0
\end{pmatrix}
\\
\gamma_1\doteq\begin{pmatrix}
	0&1\\1&0
\end{pmatrix},
\gamma_2\doteq\begin{pmatrix}
	0&-i\\i&0
\end{pmatrix}
\end{gather} 

Now let's consider the $N=2$ case, where 
\begin{align}
	\Psi_1=\frac{\gamma_1+i\gamma_2}{2}, \Psi_2=\frac{\gamma_3+i\gamma_4}{2}
\end{align}
and we also have their complex conjugates. The basis states are $\ket{0},\Psi_1^{\dagger}\ket{0},\Psi_2^{\dagger}\ket{0},\Psi_1^{\dagger}\Psi_2^{\dagger}\ket{0}$.
The matrix representations of these operators are given as follows:
\begin{gather}
\begin{split}
\Psi_1\doteq\begin{pmatrix}
		0&1&0&0\\0&0&0&0\\0&0&0&1\\0&0&0&0
	\end{pmatrix}
,\Psi_2\doteq\begin{pmatrix}
	0&0&1&0\\0&0&0&-1\\0&0&0&0\\0&0&0&0
\end{pmatrix}
,
\Psi_1^{\dagger}\doteq\begin{pmatrix}
	0&0&0&0\\1&0&0&0\\0&0&0&0\\0&0&1&0
\end{pmatrix}
,
\Psi_2^{\dagger}\doteq\begin{pmatrix}
	0&0&0&0\\0&0&0&0\\1&0&0&0\\0&-1&0&0
\end{pmatrix}
\\
\gamma_1\doteq\begin{pmatrix}
	0&1&0&0\\1&0&0&0\\0&0&0&1\\0&0&1&0
\end{pmatrix}
,\gamma_2\doteq\begin{pmatrix}
	0&-i&0&0\\i&0&0&0\\0&0&0&-i\\0&0&i&0
\end{pmatrix}
,
\gamma_3\doteq\begin{pmatrix}
	0&0&1&0\\0&0&0&-1\\1&0&0&0\\0&-1&0&0
\end{pmatrix}
,
\gamma_4\doteq\begin{pmatrix}
	0&0&-i&0\\0&0&0&i\\i&0&0&0\\0&-i&0&0
\end{pmatrix}
\end{split}
\end{gather}
We also have their commutation relations:
\begin{gather}
	[\gamma_1,\gamma_2]=\left(
	\begin{array}{cccc}
		2 i & 0 & 0 & 0 \\
		0 & -2 i & 0 & 0 \\
		0 & 0 & 2 i & 0 \\
		0 & 0 & 0 & -2 i \\
	\end{array}
	\right),
	[\gamma_2,\gamma_3]=\left(
	\begin{array}{cccc}
		0 & 0 & 0 & 2 i \\
		0 & 0 & 2 i & 0 \\
		0 & 2 i & 0 & 0 \\
		2 i & 0 & 0 & 0 \\
	\end{array}
	\right),
	[\gamma_3,\gamma_4]=\left(
	\begin{array}{cccc}
		2 i & 0 & 0 & 0 \\
		0 & 2 i & 0 & 0 \\
		0 & 0 & -2 i & 0 \\
		0 & 0 & 0 & -2 i \\
	\end{array}
	\right)
\end{gather}
We define the braiding matrices to be 
\begin{equation}
\tau(T_i)=\exp(\frac{\pi}{4}\gamma_{i+1}\gamma_i)=\frac{1}{\sqrt2}(1+\gamma_{i+1}\gamma_i).
\end{equation}
Then
\begin{gather}
	\tau(T_1)=\begin{pmatrix}
		e^{-i\pi/4}&0 &0 &0 \\
		 0&e^{i\pi/4} &0 &0 \\
		 0&0 &e^{-i\pi/4}&0 \\
		 0& 0& 0& e^{i\pi/4}
	\end{pmatrix}
,
    \tau(T_3)=\begin{pmatrix}
    	e^{-i\pi/4}& & & \\
    	&e^{-i\pi/4} & & \\
    	& &e^{i\pi/4}& \\
    	& & & e^{i\pi/4}
    \end{pmatrix}
\end{gather}
\begin{gather}
    \tau(T_2)=\frac{1}{\sqrt2}\begin{pmatrix}
    	1&0&0&-i\\
    	0&1&-i&0\\
    	0&-i&1&0\\
    	-i&0&0&1
    \end{pmatrix}.
\end{gather}

We can identify $\tau(T_1)$ as the $R-$matrix which only differs from the \eqref{eq:R12Ising} computed previously by phase rotations. Now the braid group relations also hold, for example:
\begin{equation}
    \tau(T_1)\tau(T_2)\tau(T_1)=\tau(T_2)\tau(T_1)\tau(T_2).
\end{equation}
as can be easily checked. The braiding here can be viewed as ``rotations by $\pi/2$'' in the spinor representation of $SO(4)$. For example, $\tau(T_1)=\exp(\frac14[\gamma_1,\gamma_2]*\frac\pi2)$.

\section{Solving $6$-point $\mathfrak{su}(2)_2$ KZ-equations}
It satisfies five KZ equations:
\begin{align}
	-\kappa \partial_{z_1}\psi=\left(\frac{\Omega_{12}}{z_1-z_2}+\frac{\Omega_{13}}{z_1-z_3}+\frac{\Omega_{14}}{z_1-z_4}+\frac{\Omega_{15}}{z_1-z_5}\right)\psi,\\
	-\kappa \partial_{z_2}\psi=\left(\frac{\Omega_{21}}{z_2-z_1}+\frac{\Omega_{23}}{z_2-z_3}+\frac{\Omega_{24}}{z_2-z_4}+\frac{\Omega_{25}}{z_2-z_5}\right)\psi,\\
	-\kappa \partial_{z_3}\psi=\left(\frac{\Omega_{31}}{z_3-z_1}+\frac{\Omega_{32}}{z_3-z_2}+\frac{\Omega_{34}}{z_3-z_4}+\frac{\Omega_{35}}{z_3-z_5}\right)\psi,\\
	-\kappa \partial_{z_4}\psi=\left(\frac{\Omega_{41}}{z_4-z_2}+\frac{\Omega_{42}}{z_4-z_3}+\frac{\Omega_{43}}{z_4-z_3}+\frac{\Omega_{45}}{z_4-z_5}\right)\psi,\\
	-\kappa \partial_{z_5}\psi=\left(\frac{\Omega_{51}}{z_5-z_1}+\frac{\Omega_{52}}{z_5-z_2}+\frac{\Omega_{53}}{z_5-z_3}+\frac{\Omega_{54}}{z_5-z_4}\right)\psi~.
\end{align}
We employ the following change of variables:
$$y=z_5-z_1\quad x=\frac{z_4-z_1}{z_5-z_1}\quad v=\frac{z_3-z_1}{z_4-z_1}\quad w=\frac{z_2-z_1}{z_3-z_1}\quad t=z_1+z_2+z_3+z_4+z_5$$
The inverse tranformation is
\begin{gather}
	z_1=\frac15(t-(y+yx+yxv+yxvw))\qquad z_2=xvyw+\frac15(t-(y+yx+yxv+yxvw))\qquad \nonumber\\
	z_3=xvy+\frac15(t-(y+yx+yxv+yxvw))\qquad z_4=xy+\frac15(t-(y+yx+yxv+yxvw))\qquad\nonumber\\ z_5=y+\frac15(t-(y+yx+yxv+yxvw))	
\end{gather}
Then the equations become
\begin{equation}
\begin{gathered}
	-\kappa\left(-\partial_y+\frac{x-1}{y}\partial_x+\frac{v-1}{yx}\partial_v+\frac{w-1}{yxv}\partial_w+\partial_t\right)\psi=\left(\frac{\Omega_{12}}{-yxvw}+\frac{\Omega_{13}}{-yxv}+\frac{\Omega_{14}}{-yx}+\frac{\Omega_{15}}{-y}\right)\psi\\
	-\kappa\left(\frac{1}{yxv}\partial_w+\partial_t\right)\psi=\left(\frac{\Omega_{12}}{yxvw}+\frac{\Omega_{23}}{yxvw-yxv}+\frac{\Omega_{24}}{yxvw-yx}+\frac{\Omega_{25}}{yxvw-y}\right)\psi\\
	-\kappa\left(\frac{1}{yx}\partial_v+\frac{-w}{yxv}\partial_w+\partial_t\right)\psi=\left(\frac{\Omega_{13}}{yxv}+\frac{\Omega_{23}}{yxv-yxvw}+\frac{\Omega_{34}}{yxv-yx}+\frac{\Omega_{35}}{yxv-y}\right)\psi\\
	-\kappa\left(\frac{1}{y}\partial_x+\frac{-v}{yx}\partial_v+\partial_t\right)\psi=\left(\frac{\Omega_{14}}{yx}+\frac{\Omega_{24}}{yx-yxvw}+\frac{\Omega_{34}}{yx-yxv}+\frac{\Omega_{45}}{yx-y}\right)\psi\\
	-\kappa\left(\partial_y+\frac{-x}{y}\partial_x+\partial_t\right)\psi=\left(\frac{\Omega_{15}}{y}+\frac{\Omega_{25}}{y-yxvw}+\frac{\Omega_{35}}{y-yxv}+\frac{\Omega_{45}}{y-yx}\right)\psi
\end{gathered}
\end{equation}  
We can simplify them into
\begin{equation}
\label{right kz equation}
\begin{gathered}
	-\kappa\partial_t\psi=0\\
	-\kappa\partial_y\psi=\frac{\Omega_1+\Omega_2+\Omega_3+\Omega_4}{y}\psi\\
	-\kappa\partial_w\psi=\left(\frac{\Omega_{12}}{w}+\frac{\Omega_{23}}{w-1}+\frac{v\Omega_{24}}{vw-1}+\frac{xv\Omega_{25}}{xvw-1}\right)\psi\\
	-\kappa\partial_v\psi=\left(\frac{\Omega_{12}+\Omega_{13}+\Omega_{23}}{v}+\frac{\Omega_{34}}{v-1}+\frac{w\Omega_{24}}{vw-1}+\frac{x\Omega_{35}}{xv-1}+\frac{wx\Omega_{25}}{xvw-1}\right)\psi\\
	-\kappa\partial_x\psi=\left(\frac{\Omega_{12}+\Omega_{13}+\Omega_{14}+\Omega_{23}+\Omega_{24}+\Omega_{34}}{x}+\frac{\Omega_{45}}{x-1}+\frac{v\Omega_{35}}{xv-1}+\frac{wv\Omega_{25}}{xvw-1}\right)\psi
	\end{gathered}.
\end{equation}
$\psi$ lives in the tensor product vector space $(\frac12)^6$, we can write it as $\psi=F_1I_1+F_2I_2+F_3I_3+F_4I_4+F_5I_5$, where $I_j$-symbols are defined as:
\begin{align}
\label{6pbasis}
\begin{split}
	I_1=&\frac{1}{2\sqrt2}(\doa\upa\doa\upa\doa\upa-\upa\doa\doa\upa\doa\upa-\doa\upa\upa\doa\doa\upa+\upa\doa\upa\doa\doa\upa-\doa\upa\doa\upa\upa\doa+\upa\doa\doa\upa\upa\doa+\doa\upa\upa\doa\upa\doa-\upa\doa\upa\doa\upa\doa)\\
	&=(\frac12,0,\frac12,0,\frac12,0);\\
	I_2=&\frac{1}{\sqrt 6}(\doa\upa\doa\doa\upa\upa-\upa\doa\doa\doa\upa\upa+\doa\upa\upa\upa\doa\doa-\upa\doa\upa\upa\doa\doa)+\\
	&\frac{1}{2 \sqrt{6}}(\upa\doa\doa\upa\doa\upa-\doa\upa\upa\doa\doa\upa+\upa\doa\upa\doa\doa\upa-\doa\upa\doa\upa\upa\doa+\upa\doa\doa\upa\upa\doa-\doa\upa\upa\doa\upa\doa+\upa\doa\upa\doa\upa\doa-\doa\upa\doa\upa\doa\upa)
	\\
	&=(\frac12,0,\frac12,1,\frac12,0);\\
	I_3=&\frac{\doa\doa\upa\upa\doa\upa}{\sqrt{6}}-\frac{\doa\upa\doa\upa\doa\upa}{2 \sqrt{6}}-\frac{\upa\doa\doa\upa\doa\upa}{2 \sqrt{6}}-
	\frac{\doa\upa\upa\doa\doa\upa}{2 \sqrt{6}}-\frac{\upa\doa\upa\doa\doa\upa}{2 \sqrt{6}}+\frac{\upa\upa\doa\doa\doa\upa}{\sqrt{6}}-\frac{\doa\doa\upa\upa\upa\doa}{\sqrt{6}}\\
	&+\frac{\doa\upa\doa\upa\upa\doa}{2 \sqrt{6}}+\frac{\upa\doa\doa\upa\upa\doa}{2 \sqrt{6}}+\frac{\doa\upa\upa\doa\upa\doa}{2 \sqrt{6}}+\frac{\upa\doa\upa\doa\upa\doa}{2 \sqrt{6}}
	-\frac{\upa\upa\doa\doa\upa\doa}{\sqrt{6}}\\
	&=(\frac12,1,\frac12,0,\frac12,0);\\
	I_4=&\frac{\sqrt{2}}{3}\doa\doa\upa\doa\upa\upa-\frac{\doa\upa\doa\doa\upa\doa}{3 \sqrt{2}}-\frac{\upa\doa\doa\doa\upa\upa}{3 \sqrt{2}}-\frac{\doa\doa\upa\upa\doa\upa}{3 \sqrt{2}}+\frac{\doa\upa\doa\upa\doa\upa}{6 \sqrt{2}}+\frac{\upa\doa\doa\upa\doa\upa}{6 \sqrt{2}}\\
	&-\frac{\doa\upa\upa\doa\doa\upa}{6 \sqrt{2}}-\frac{\upa\doa\upa\doa\doa\upa}{6 \sqrt{2}}+\frac{\upa\upa\doa\doa\doa\upa}{3 \sqrt{2}}
	-\frac{\doa\doa\upa\upa\upa\doa}{3 \sqrt{2}}+\frac{\doa\upa\doa\upa\upa\doa}{6 \sqrt{2}}+\frac{\upa\doa\doa\upa\upa\doa}{6 \sqrt{2}}-\frac{\doa\upa\upa\doa\upa\doa}{6
		\sqrt{2}}\\
	&-\frac{\upa\doa\upa\doa\upa\doa}{6 \sqrt{2}}+\frac{\upa\upa\doa\doa\upa\doa}{3 \sqrt{2}}+\frac{\doa\upa\upa\upa\doa\doa}{3 \sqrt{2}}
	+\frac{\upa\doa\upa\upa\doa\doa}{3 \sqrt{2}}-\frac{\sqrt{2}}{3}\upa\upa\doa\upa\doa\doa\\
	&=(\frac12,1,\frac12,1,\frac12,0);\\
	I_5=&-\frac{1}{2} \doa\doa\doa\upa\upa\upa+\frac{1}{6} \doa\doa\upa\doa\upa\upa+\frac{1}{6} \doa\upa\doa\doa\upa\doa+\frac{1}{6} \upa\doa\doa\doa\upa\upa+\frac{1}{6} \doa\doa\upa\upa\doa\upa+\frac{1}{6} \doa\upa\doa\upa\doa\upa+\frac{1}{6} \upa\doa\doa\upa\doa\upa\\
	&-\frac{1}{6} \doa\upa\upa\doa\doa\upa-\frac{1}{6} \upa\doa\upa\doa\doa\upa
	-\frac{1}{6} \upa\upa\doa\doa\doa\upa+\frac{1}{6} \doa\doa\upa\upa\upa\doa+\frac{1}{6} \doa\upa\doa\upa\upa\doa+\frac{1}{6}
	\upa\doa\doa\upa\upa\doa-\frac{1}{6} \doa\upa\upa\doa\upa\doa\\
	&-\frac{1}{6} \upa\doa\upa\doa\upa\doa-\frac{1}{6}\upa\upa\doa\doa\upa\doa-\frac{1}{6} \doa\upa\upa\upa\doa\doa-\frac{1}{6} \upa\doa\upa\upa\doa\doa-\frac{1}{6} \upa\upa\doa\upa\doa\doa+\frac{1}{2} \upa\upa\upa\doa\doa\doa\\
	&=(\frac12,1,\frac32,1,\frac12,0).
	\end{split}
\end{align}
These basis states correspond to the fusion channel
\begin{equation}
\label{fusion channel 6}
	(((((\mathcal O_1\otimes\mathcal O_2)\otimes\mathcal O_3)\otimes\mathcal O_4)\otimes\mathcal O_5)\otimes\mathcal O_6),
\end{equation}
which can be identified with the set of trees:
\begin{center}
\scalebox{1.2}
{
\begin{tikzpicture}
		
		\draw[black, thick] (2,1) -- (5.5,-2.5);
		\draw (2.1,1) node[anchor=west]{$ {\frac12}$};
		\draw[black, thick] (3.4,1)node[anchor=west]{$ {\frac12}$}--(2.7,0.3)node[anchor=west] {$ {a}$};
		\draw[black, thick] (4.5,1)node[anchor=west]{$ {\frac12}$}--(3.25,-0.25)node[anchor=west] {$ {b}$};
		\draw[black, thick] (5.6,1)node[anchor=west]{$ {\frac12}$}--(3.8,-0.8)node[anchor=west] {$ {c}$};
		\draw[black, thick] (6.7,1)node[anchor=west]{$ {\frac12}$}--(4.35,-1.35)node[anchor=west] {$ {d}$};
		\draw[black, thick] (7.8,1)node[anchor=west]{$ {\frac12}$}--(4.9,-1.9)node[anchor=west] {$ {e}$};
	    \draw(8.9,-1.9)node[anchor=west]{$\text{,}$};
	\end{tikzpicture}
	}
	\end{center}
where the ordered set $\{\frac{1}{2},a,b,c,d,e\}$ denote different fusion possibilities. For example, $I_1$ corresponds to the set $\{ \frac12,0,\frac12,0,\frac12,0\}$.

There is a new feature we should take into consideration in the $6$-point case. The quantum Clebsch-Gordon rule would put restrictions on the internal spin of fusion trees \cite{kohno2002conformal}. Therefore, $I_5$ should be ruled out as a valid tree. We will see how this mechanism works in detail soon. 
The action of $\Omega_{ij}$ is the same as in the 4-point case. Note that each $I_i$ is an eigenvector of $\Omega_{i}$s. 
We have
\begin{align}
\nonumber
\large
	\renewcommand{\arraystretch}{1.2}
	\begin{array}{c|c|c|c}
		  & I_1 & I_2 & I_3 \\
		  \hline
		\Omega _{12} & -\frac{1}{2} \left(3 I_1\right) & -\frac{1}{2} \left(3 I_2\right) & \frac{I_3}{2} \\
		\hline
		\Omega _{13} & -\frac{1}{2} \sqrt{3} I_3 & -\frac{1}{2} \sqrt{3} I_4 & -\frac{1}{2} \sqrt{3} I_1-I_3 \\
		\hline
		\Omega _{14} & \frac{\sqrt{3} I_3}{2} & -\frac{I_4-2 \sqrt{2} I_5}{2 \sqrt{3}} & \frac{\sqrt{3} I_1}{2}-I_3 \\
		\hline
		\Omega _{15} & \frac{1}{2} \left(I_4+\sqrt{2} I_5\right) & \frac{I_3}{2}+\frac{I_4}{\sqrt{3}}-\frac{I_5}{\sqrt{6}} & \frac{1}{6} \left(3 I_2-2 \sqrt{3} I_4+\sqrt{6} I_5\right) \\
		\hline
		\Omega _{23} & \frac{\sqrt{3} I_3}{2} & \frac{\sqrt{3} I_4}{2} & \frac{\sqrt{3} I_1}{2}-I_3 \\
		\hline
		\Omega _{24} & -\frac{1}{2} \sqrt{3} I_3 & \frac{I_4-2 \sqrt{2} I_5}{2 \sqrt{3}} & -\frac{1}{2} \sqrt{3} I_1-I_3 \\
		\hline
		\Omega _{25} & -\frac{I_4}{2}-\frac{I_5}{\sqrt{2}} & \frac{1}{6} \left(-3 I_3-2 \sqrt{3} I_4+\sqrt{6} I_5\right) & \frac{1}{6} \left(-3 I_2-2 \sqrt{3} I_4+\sqrt{6} I_5\right) \\
		\hline
		\Omega _{34} & -\frac{1}{2} \left(3 I_1\right) & \frac{I_2}{2} & \frac{I_3}{2} \\
		\hline
		\Omega _{35} & -\frac{1}{2} \sqrt{3} I_2 & -\frac{1}{2} \sqrt{3} I_1-I_2 & \frac{I_4-2 \sqrt{2} I_5}{2 \sqrt{3}} \\
		\hline
		\Omega _{45} & \frac{\sqrt{3} I_2}{2} & \frac{\sqrt{3} I_1}{2}-I_2 & \frac{\sqrt{3} I_4}{2} \\
	\end{array}
\end{align}
\begin{equation}
	\large
	\renewcommand{\arraystretch}{1.2}
	\begin{array}{c|c}
	 & I_4  \\
	\hline
	\Omega _{12} & \frac{I_4}{2}  \\
	\hline
	\Omega _{13} & -\frac{1}{2} \sqrt{3} I_2-I_4 \\
	\hline
	\Omega _{14} & \frac{1}{6} \left(2 \left(I_4+\sqrt{2} I_5\right)-\sqrt{3} I_2\right) \\
	\hline
	\Omega _{15} & \frac{1}{6} \left(3 I_1+2 \sqrt{3} I_2-2 \sqrt{3} I_3-4 I_4-\sqrt{2} I_5\right) \\
	\hline
	\Omega _{23} & \frac{\sqrt{3} I_2}{2}-I_4 \\
	\hline
	\Omega _{24} & \frac{1}{6} \left(\sqrt{3} I_2+2 \left(I_4+\sqrt{2} I_5\right)\right)  \\
	\hline
	\Omega _{25} & \frac{1}{6} \left(-3 I_1-2 \sqrt{3} I_2-2 \sqrt{3} I_3-4 I_4-\sqrt{2} I_5\right)  \\
	\hline
	\Omega _{34} & \frac{1}{6} \left(-I_4-4 \sqrt{2} I_5\right) \\
	\hline
	\Omega _{35} & \frac{1}{6} \left(\sqrt{3} I_3+2 \left(I_4+\sqrt{2} I_5\right)\right)  \\
	\hline
	\Omega _{45} & \frac{\sqrt{3} I_3}{2}-I_4 \\
\end{array}
\end{equation}
\begin{equation}
\nonumber
	\large
	\renewcommand{\arraystretch}{1.2}
	\begin{array}{c|c}
	  & I_5 \\
	\hline
	\Omega _{12}  & \frac{I_5}{2} \\
	\hline
	\Omega _{13}  & \frac{I_5}{2} \\
	\hline
	\Omega _{14} & \frac{1}{6} \left(2 \sqrt{6} I_2+2 \sqrt{2} I_4-5 I_5\right) \\
	\hline
	\Omega _{15}  & \frac{1}{6} \left(3 \sqrt{2} I_1-\sqrt{6} I_2+\sqrt{6} I_3-\sqrt{2} I_4-5 I_5\right) \\
	\hline
	\Omega _{23} & \frac{I_5}{2} \\
	\hline
	\Omega _{24} & \frac{1}{6} \left(-2 \sqrt{6} I_2+2 \sqrt{2} I_4-5 I_5\right) \\
	\hline
	\Omega _{25}& \frac{1}{6} \left(-3 \sqrt{2} I_1+\sqrt{6} I_2+\sqrt{6} I_3-\sqrt{2} I_4-5 I_5\right) \\
	\hline
	\Omega _{34}  & \frac{1}{6} \left(-4 \sqrt{2} I_4-5 I_5\right) \\
	\hline
	\Omega _{35}  & \frac{1}{6} \left(-2 \sqrt{6} I_3+2 \sqrt{2} I_4-5 I_5\right) \\
	\hline
	\Omega _{45} & \frac{I_5}{2} \\
\end{array}
\end{equation}
Now we can plug all these actions into \eqref{right kz equation}. The result can be written compactly using differential forms.
Let $\vec{F}=(F_1,F_2,F_3,F_4,F_5)^T$, then
\begin{equation}
\label{kz6}
	d\vec{F}=\partial_x\vec{F}dx+\partial_v\vec{F}dv+\partial_w\vec{F}dw=\frac1\kappa\Omega \vec{F},
\end{equation}
where $\Omega$ is the $1$-form valued matrix:
$$\Omega=A_xdx+A_vdv+A_wdw.$$
Now we focus on one component of the one-form equation $\partial_wF=A_wF$, sending $v,x\rightarrow 0$. It is a fuchsian differential equation in matrix form:
\begin{align}
	\partial_w\vec{F}=\frac1\kappa\left(
	\begin{array}{ccccc}
		\frac{3}{2 w} & 0 & -\frac{\sqrt{3}}{2 (w-1)} & 0&0 \\
		0 & \frac{3}{2 w} & 0 & -\frac{\sqrt{3}}{2 (w-1)}&0 \\
		-\frac{\sqrt{3}}{2 (w-1)} & 0 & -\frac{1}{2 w}+\frac{1}{w-1} & 0&0 \\
		0 & -\frac{\sqrt{3}}{2 (w-1)} & 0 & -\frac{1}{2 w}+\frac{1}{w-1}&0 \\
		0&0&0&0&-\frac{1}{2(w-1)}-\frac{1}{2w}
	\end{array}
	\right)\vec{F}.
\end{align}
We can see that $F_1,F_3$ and $F_2,F_4$ decouple from each other. And they satisfy identical equations. Letting $\kappa=4$, we can write the equations for $F_1\&F_3$ explicitly:
\begin{align}
	\label{eqn original}
	\begin{split}
	\partial_w F_1&=\frac3{8w}F_1-\frac{\sqrt3}{8(w-1)}F_3,\\
	\partial_w F_3&=\frac{\sqrt3}{8(1-w)}F_1-(\frac{1}{8w}-\frac{1}{4(w-1)})F_3.
\end{split}
\end{align}

Using an ansatz $F_1(w)=w^{3/8}(1-w)^{-1/8}f(w)$, we obtain a hypergeometric equation: 
$$w(1-w)f''(w)+(\frac12-w) f'(w)+\frac1{16}f(w)=0.$$
The parameters are $a=\frac14,b=-\frac14,c=\frac12$.

Now we have $2$ linearly independent solutions of \eqref{eqn original}
\begin{gather}
	F_1^{+}=w^{\frac{3}{8}}(1-w)^{-\frac{1}{8}}u_1,\\	
	F_1^{-}=w^{\frac{3}{8}}(1-w)^{-\frac{1}{8}}u_2,
\end{gather}
together with the corresponding $F_3^{\pm}$.
Since the $F_2\&F_4$ satisfy the same equation set as $F_1\&F_3$, the solutions are also the same. We use $F_2^{\pm}$ and $F_4^{\pm}$ to denote the corresponding solutions.
$F_5$ can be solved directly: $F_5=w^{-1/8}(1-w)^{-1/8}$.
So in total there are $5$ linearly independent solutions of the original Fuchsian system $\partial_w\vec F=A_w\vec F$ (denoted as curly $\mathcal{F}_i$s):
\begin{gather}
\mathcal{F}_1= \begin{pmatrix}
F_1^{+}\\
0\\
F_3^{+}\\
0\\
0
\end{pmatrix},
\mathcal{F}_2=\begin{pmatrix}
F_1^{-}\\
0\\
F_3^{-}\\
0\\
0
\end{pmatrix},
\mathcal{F}_3=\begin{pmatrix}
0\\
F_2^{+}\\
0\\
F_4^{+}\\
0
\end{pmatrix},
\mathcal{F}_4=\begin{pmatrix}
0\\
F_2^{-}\\
0\\
F_4^{-}\\
0
\end{pmatrix},
\mathcal{F}_5=
\begin{pmatrix}
0\\0\\0\\0\\F_5
\end{pmatrix}.
\end{gather}
The notation means $\mathcal{F}_1=F^+_1I_1+F^+_3I_3$ and so on. Following the same logic as in the  $4$-point case, we can associate these blocks to local basis states. $\mathcal{F}_5$ is the one associated to $I_5$, which is forbidden by the quantum Clebsch-Gordon rules. So we can ignore this solution and the space of the conformal blocks are spanned by $\mathcal{F}_{1,2,3,4}$.

Sending $w,x\rightarrow0$ in \eqref{kz6}, we find that $\partial_v F=A_vF$ gives us a simple set of equations,
\begin{equation}
\begin{gathered}
	F_1'(v)=\frac{1}{4} \left(\frac{3}{2 v}+\frac{3}{2 (v-1)}\right) F_1(v),\\
	F_2'(v)=\frac{1}{4} \left(-\frac{1}{2 (v-1)}+\frac{3}{2 v}\right) F_2(v),\\
	F_3'(v)=\frac{1}{4} \left(-\frac{1}{2 (v-1)}+\frac{3}{2 v}\right) F_3(v),\\
	F_4'(v)=\frac{1}{4} \left(\frac{3}{2 v}+\frac{1}{6 (v-1)}\right) F_4(v)+\frac{F_5(v)}{3 \sqrt{2} (v-1)},\\
	F_5'(v)=\frac{1}{4} \left(-\frac{3}{2 v}+\frac{5}{6 (v-1)}\right) F_5(v)+\frac{F_4(v)}{3 \sqrt{2} (v-1)}.
	\end{gathered}
\end{equation}
The first three equations are easy to solve. For example, $F_1=v^{3/8}(1-v)^{3/8}$. The coupled equation of $F_4$ and $F_5$ can not be transformed into a hypergeometric one, but using the ansatz $F_4=v^{5/8}(1-v)^{-1/8}f(v)$, we can write out the equation for $f(v)$,
\begin{equation}
	v(1-v)f''(v)+\frac{1}{4}(7v-5)f'(v)=0.
\end{equation}
This is in fact a first order equation, the solution is 
$$f'(v)=0\qquad or\qquad f'(v)=v^{-4/5}(1-v)^{-1/2},$$
By integration,
$$f(v)=const.\qquad or\qquad f(v)=\int v^{-4/5}(1-v)^{-1/2}dv=\sqrt{1-v}\prescript{}{2}F_1(\frac12,\frac54,\frac32,1-v).$$
So the original first order system has the following 2 solutions:
\begin{equation}
	\begin{pmatrix}
		F_4^+=v^{\frac58}(1-v)^{-1/8}\\
		F_5^+=DF_4^+
	\end{pmatrix},
\quad
\begin{pmatrix}
	F_4^-=v^{\frac58}(1-v)^{\frac38}\prescript{}{2}F_1(\frac12,\frac54,\frac32,1-v)\\
	F_5^-=DF_4^-
\end{pmatrix},
\end{equation}
where $D$ is a linear operator: $D\equiv \frac{1}{4\sqrt2v}(24v(v-1)\partial_v+(9-10v))$.

Since $v=\frac{z_3-z_1}{z_4-z_1}$, $B_{34}$ is associated to the monodromy around $v=1$. 
Again, we need to switch to a new fusion channel. In this case we use 
\begin{gather}
    I'_1=I_1,\quad I'_2=I_2, \quad I'_3=I_3, \nonumber\\
    I'_4=-\frac{1}{\sqrt3}I_4-\frac{\sqrt2}{\sqrt3}I_5,\quad I'_5=-\frac{\sqrt2}{\sqrt3}I_4+\frac{1}{\sqrt3}I_5
\end{gather}

Then 
\begin{gather}
    F_4^+I_4+F_5^+I_5=F^{\prime +}_4I'_4+F^{\prime +}_5I'_5\\
    F_4^-I_4+F_5^-I_5=F^{\prime -}_4I'_4+F^{\prime -}_5I'_5
\end{gather}
with scaling 
\begin{equation}
    \left\{
	\begin{aligned}
		F_4^{\prime+}&\sim\frac{\sqrt3}{2}(1-v)^{7/8} \\
		F_5^{\prime+}&\sim -\frac{\sqrt3}{\sqrt2}(1-v)^{-1/8}
	\end{aligned}
	\right.\qquad
	\left\{
	\begin{aligned}
		F_4^{\prime-}&\sim -\sqrt3 (1-v)^{3/8}\\
		F_5^{\prime-}&\sim \frac{1}{\sqrt6}(1-v)^{11/8}
	\end{aligned}
	\right.
\end{equation}
The $+$ blocks corresponds to $I'_5$, which is forbidden by the quantum Clebsch-Gordon rule. And all the blocks are eigenfunctions under the monodromy, so we don't need an $F$-move. Finally we have
$$B_{34}=R_{34}=\begin{pmatrix}
	-e^{-\frac38\pi i}&&&\\
	&e^{\frac18\pi i}&&\\
	&&e^{\frac18\pi i}&\\
	&&&-e^{-\frac{3}{8}\pi i}
\end{pmatrix}.$$
We can check the braiding relation
\begin{equation}
    B_{34}B_{23}B_{34}=B_{23}B_{34}B_{23}
\end{equation}
Next, let's have a look at $\partial_x F=A_xF$. Sending $v,w\rightarrow0$, we get
\begin{align}
	\partial_x\vec{F}=\frac1\kappa
	\left(
	\begin{array}{ccccc}
		\frac{3}{4 x} & -\frac{\sqrt{3}}{8 (x-1)} & 0 & 0 & 0 \\
		-\frac{\sqrt{3}}{8 (x-1)} & \frac{1}{4} \left(\frac{1}{x}+\frac{1}{x-1}\right) & 0 & 0 & 0 \\
		0 & 0 & \frac{3}{4 x} & -\frac{\sqrt{3}}{8 (x-1)} & 0 \\
		0 & 0 & -\frac{\sqrt{3}}{8 (x-1)} & \frac{1}{4} \left(\frac{1}{x}+\frac{1}{x-1}\right) & 0 \\
		0 & 0 & 0 & 0 & \frac{1}{4} \left(\frac{1}{x}-\frac{1}{2 (x-1)}\right) \\
	\end{array}
	\right)
	\vec{F}.
\end{align}
$F_5$ decouples naturally. 
The computation of $B_{45}$ goes along the same lines as detailed above. 
The basis transformation we need is :
\begin{gather}
P_1=\frac14(I_1-\sqrt3I_2-\sqrt3I_3+3I_4);\nonumber\\
P_2=\frac14(\sqrt3I_1+I_2-3I_3-\sqrt3I_4);\nonumber\\
P_3=\frac14(\sqrt3I_1-3I_2+I_3-\sqrt3I_4);\nonumber\\
P_4=\frac14(3I_1+\sqrt3I_2+\sqrt3I_3+I_4);\nonumber\\
P_5=-I_5.
\end{gather}
The $R_{45}$ matrix is
\begin{align}
R_{45}=\begin{pmatrix}
	e^{\frac18\pi i}&&&\\
	&-e^{-\frac38\pi i}&&\\
	&&e^{\frac18\pi i}&\\
	&&&-e^{-\frac38\pi i}
\end{pmatrix},
\end{align}
and the $F$-move is the same as in the previous case. So finally we obtain
\begin{align}
B_{45}=FR_{45}F^{-1}=\frac{(-1)^{\frac58}}{2}\left(
\begin{array}{cccc}
1-i& -1-i & 0 & 0 \\
	-1-i &1-i & 0 & 0 \\
	0 & 0 & 1-i & -1-i\\
	0 & 0 & -1-i & 1-i \\
\end{array}
\right).
\end{align}
It is easy to check that the braid group relation is also satisfied:
\begin{equation}
	B_{45}B_{34}B_{45}=B_{34}B_{45}B_{34}.
\end{equation}
\section{Spin basis in the 4 point Fibonacci anyon }
If we use $\ket{0,\pm1}$ to denote the different spin states in the $j=1$ rep, then the $3$ basis states in the final tensored space are:
\begin{align}
Q_1=&-\frac{1}{3} \ket1_2 \ket0_3 \ket0_4 \ket{-1}_1-\frac{1}{3} \ket1_1 \ket0_3 \ket0_4 \ket{-1}_2-\frac{1}{3} \ket1_4 \ket0_1 \ket0_2 \ket{-1}_3-\frac{1}{3} \ket1_3 \ket0_1 \ket0_2 \ket{-1}_4\nonumber\\
&+\frac{1}{3} \ket1_2 \ket1_4 \ket{-1}_1 \ket{-1}_3+\frac{1}{3} \ket1_1 \ket1_4 \ket{-1}_2 \ket{-1}_3+\frac{1}{3} \ket1_2 \ket1_3 \ket{-1}_1 \ket{-1}_4\nonumber\\
&+\frac{1}{3} \ket1_1 \ket1_3 \ket{-1}_2 \ket{-1}_4+\frac{1}{3} \ket0_1 \ket0_2 \ket0_3 \ket0_4
=(1,0,1,0);\\
Q_2=&\frac{\ket1_4 \ket0_2 \ket0_3 \ket{-1}_1}{2 \sqrt{3}}-\frac{\ket1_3 \ket0_2 \ket0_4 \ket{-1}_1}{2 \sqrt{3}}+\frac{\ket1_3 \ket0_1 \ket0_4 \ket{-1}_2}{2 \sqrt{3}}+\frac{\ket1_2 \ket0_1 \ket0_4 \ket{-1}_3}{2 \sqrt{3}}\nonumber\\
&+\frac{\ket1_1 \ket0_2 \ket0_3 \ket{-1}_4}{2 \sqrt{3}}-\frac{\ket1_4 \ket0_1 \ket0_3 \ket{-1}_2}{2 \sqrt{3}}-\frac{\ket1_1 \ket0_2 \ket0_4 \ket{-1}_3}{2 \sqrt{3}}-\frac{\ket1_2 \ket0_1 \ket0_3 \ket{-1}_4}{2 \sqrt{3}}\nonumber\\
&+\frac{\ket1_2 \ket1_3 \ket{-1}_4 \ket{-1}_1}{2 \sqrt{3}}-\frac{\ket1_2 \ket1_4 \ket{-1}_3 \ket{-1}_1}{2 \sqrt{3}}+\frac{\ket1_1 \ket1_4 \ket{-1}_2 \ket{-1}_3}{2 \sqrt{3}}\nonumber\\&-\frac{\ket1_1 \ket1_3 \ket{-1}_2 \ket{-1}_4}{2 \sqrt{3}}
=(1,1,1,0);\\
Q_3=&\frac{\ket1_2 \ket0_3 \ket0_4 \ket{-1}_1}{3 \sqrt{5}}+\frac{\ket1_1 \ket0_3 \ket0_4 \ket{-1}_2}{3 \sqrt{5}}-\frac{\ket1_3 \ket0_2 \ket0_4 \ket{-1}_1}{2 \sqrt{5}}-\frac{\ket1_3 \ket0_1 \ket0_4 \ket{-1}_2}{2 \sqrt{5}}\nonumber\\
&-\frac{\ket1_2 \ket0_1 \ket0_4 \ket{-1}_3}{2 \sqrt{5}}-\frac{\ket1_1 \ket0_2 \ket0_4 \ket{-1}_3}{2 \sqrt{5}}+\frac{\ket1_4 \ket0_1 \ket0_2 \ket{-1}_3}{3 \sqrt{5}}+\frac{\ket1_3 \ket0_1 \ket0_2 \ket{-1}_4}{3 \sqrt{5}}\nonumber\\
&-\frac{\ket1_4 \ket0_2 \ket0_3 \ket{-1}_1}{2 \sqrt{5}}-\frac{\ket1_4 \ket0_1 \ket0_3 \ket{-1}_2}{2 \sqrt{5}}-\frac{\ket1_2 \ket0_1 \ket0_3 \ket{-1}_4}{2 \sqrt{5}}-\frac{\ket1_1 \ket0_2 \ket0_3 \ket{-1}_4}{2 \sqrt{5}}\nonumber\\
&+\frac{\ket1_3 \ket1_4 \ket{-1}_1 \ket{-1}_2}{\sqrt{5}}+\frac{\ket1_2 \ket1_4 \ket{-1}_1 \ket{-1}_3}{6 \sqrt{5}}+\frac{\ket1_1 \ket1_4 \ket{-1}_2 \ket{-1}_3}{6 \sqrt{5}}\nonumber\\&+\frac{\ket1_2 \ket1_3 \ket{-1}_1 \ket{-1}_4}{6 \sqrt{5}}+\frac{\ket1_1 \ket1_3 \ket{-1}_2 \ket{-1}_4}{6 \sqrt{5}}
+\frac{\ket1_1 \ket1_2 \ket{-1}_3 \ket{-1}_4}{\sqrt{5}}\nonumber\\&+\frac{2 \ket0_1 \ket0_2 \ket0_3 \ket0_4}{3 \sqrt{5}}
=(1,2,1,0),
\end{align}
where the subscripts label the order of the tensored vector spaces. Note here that $Q_3$ is the one with higher intermediate spin than the allowed value imposed by the quantum Clebsch-Gordon rule.

Actions of $J$'s on spin-$1$ states are:
\begin{gather*}
	J_z\ket{1}=\ket{1},\quad J_z\ket{0}=0,\quad J_z\ket{-1}=-\ket{-1}\\
	J_x\ket{1}=\frac{\sqrt2}{2}\ket{0},\quad J_x\ket{0}=\frac{\sqrt2}{2}(\ket{1}+\ket{-1}),\quad J_x\ket{-1}=\frac{\sqrt2}{2}\ket{0}\\
	J_y\ket{1}=\frac{\sqrt2i}{2}\ket{0},\quad J_y\ket{0}=-\frac{\sqrt2i}{2}(\ket{1}-\ket{-1}),\quad
	J_y\ket{-1}=-\frac{\sqrt2i}{2}\ket{0}
\end{gather*}
Actions of $\Omega$ are:
\begin{equation}
\begin{gathered}
	\ket{1}\ket{1}\rightarrow2\ket{1}\ket{1}\quad\ket{1}\ket{0}\rightarrow2\ket{0}\ket{1}\quad\ket{1}\ket{-1}\rightarrow2(\ket{0}\ket{0}-\ket{1}\ket{-1})\\
	\ket{0}\ket{0}\rightarrow2(\ket{1}\ket{-1}+\ket{-1}\ket{1})\quad\ket{0}\ket{-1}\rightarrow2\ket{-1}\ket{0}\quad\ket{-1}\ket{-1}\rightarrow2\ket{-1}\ket{-1}
	\end{gathered}
\end{equation}
\begin{equation}
	\large
	\renewcommand{\arraystretch}{1.2}
	\begin{array}{c|c|c|c}
		\text{} & Q_1 & Q_2 & Q_3 \\\hline
		\Omega _{12} & -4 Q_1 & -2 Q_2 & 2 Q_3 \\\hline
		\Omega _{13} & -\frac{4 Q_2}{\sqrt{3}} & -\frac{4 Q_1}{\sqrt{3}}-Q_2-\sqrt{\frac{5}{3}} Q_3 & -\sqrt{\frac{5}{3}} Q_2-3 Q_3 \\\hline
		\Omega _{14} & \frac{4 Q_2}{\sqrt{3}} & \frac{4 Q_1}{\sqrt{3}}-Q_2+\sqrt{\frac{5}{3}} Q_3 & \sqrt{\frac{5}{3}} Q_2-3 Q_3 \\\hline
		\Omega _{23} & \frac{4 Q_2}{\sqrt{3}} & \frac{4 Q_1}{\sqrt{3}}-Q_2+\sqrt{\frac{5}{3}} Q_3 & \sqrt{\frac{5}{3}} Q_2-3 Q_3 \\\hline
		\Omega _{24} & -\frac{4 Q_2}{\sqrt{3}} & -\frac{4 Q_1}{\sqrt{3}}-Q_2-\sqrt{\frac{5}{3}} Q_3 & -\sqrt{\frac{5}{3}} Q_2-3 Q_3 \\\hline
		\Omega _{34} & -4 Q_1 & -2 Q_2 & 2 Q_3 \\\hline
	\end{array}
\end{equation}
For transformed basis, we have
\begin{equation}
\label{tableQprime}
\begin{array}{c|c|c|c}
		\text{} & Q'_1 & Q'_2 & Q'_3 \\\hline
		\Omega _{12} & \frac{4 Q'_2}{\sqrt{3}} & \frac{4 Q'_1}{\sqrt{3}}-Q'_2+\sqrt{\frac{5}{3}} Q'_3 & \sqrt{\frac{5}{3}} Q'_2-3 Q'_3 \\\hline
		\Omega _{13} & -\frac{4 Q'_2}{\sqrt{3}} & -\frac{4 Q'_1}{\sqrt{3}}-Q'_2-\sqrt{\frac{5}{3}} Q'_3 & -\sqrt{\frac{5}{3}} Q'_2-3 Q'_3 \\\hline
		\Omega _{14} & -4 Q'_1 & -2 Q'_2 & 2 Q'_3 \\\hline
		\Omega _{23} & -4 Q'_1 & -2 Q'_2 & 2 Q'_3 \\\hline
		\Omega _{24} & -\frac{4 Q'_2}{\sqrt{3}} & -\frac{4 Q'_1}{\sqrt{3}}-Q'_2-\sqrt{\frac{5}{3}} Q'_3 & -\sqrt{\frac{5}{3}} Q'_2-3 Q'_3 \\\hline
		\Omega _{34} & \frac{4 Q'_2}{\sqrt{3}} & \frac{4 Q'_1}{\sqrt{3}}-Q'_2+\sqrt{\frac{5}{3}} Q'_3 & \sqrt{\frac{5}{3}} Q'_2-3 Q'_3 \\\hline
	\end{array}
\end{equation}
\bibliographystyle{JHEP}     
 {\small{\bibliography{references}}}
 
\end{document}